\documentclass[12pt,a4paper,english,titlepage]{article}
\usepackage[T1]{fontenc}
\usepackage[latin9]{inputenc}
\usepackage{verbatim}
\usepackage{float}
\usepackage{units}
\usepackage{amsmath}
\usepackage{amssymb}
\usepackage{graphicx}
\PassOptionsToPackage{normalem}{ulem}
\usepackage{ulem}

\makeatletter

\newcommand{\lyxmathsym}[1]{\ifmmode\begingroup\def\b@ld{bold}
  \text{\ifx\math@version\b@ld\bfseries\fi#1}\endgroup\else#1\fi}

\providecommand{\tabularnewline}{\\}

\newenvironment{lyxcode}
	{\par\begin{list}{}{
		\setlength{\rightmargin}{\leftmargin}
		\setlength{\listparindent}{0pt}
		\raggedright
		\setlength{\itemsep}{0pt}
		\setlength{\parsep}{0pt}
		\normalfont\ttfamily}%
	 \item[]}
	{\end{list}}

\usepackage{eurosym}
\usepackage{rotating}
\usepackage{amsfonts}
\usepackage{chicago}

\usepackage{color}
\usepackage{lscape}
\usepackage{ulem}
\usepackage{multirow}
\usepackage{bbm}

\newtheorem{theorem}{Theorem}

\newtheorem{axiom}{Axiom}

\newenvironment{proof}[1][Proof]{\textbf{#1.} }{\ \rule{0.5em}{0.5em}}

\makeatother

\usepackage{babel}
\begin{document}
\begin{titlepage}

\bigskip{}
 \bigskip{}

\begin{center}
{\LARGE{}{}\textbf{\LARGE{}Mobility and Mobility Measures}{\LARGE{}}}{\LARGE\par}
\par\end{center}

\begin{center}
{\LARGE{}\medskip{}
}{\LARGE\par}
\par\end{center}

\begin{center}
{\LARGE{}\medskip{}
}{\LARGE\par}
\par\end{center}

\begin{center}
by
\par\end{center}

\begin{center}
{\LARGE{}\medskip{}
 }{\Large{}{}\textbf{\Large{}Frank A. Cowell}{\Large{}}}{\Large\par}
\par\end{center}

\begin{center}
{\Large{}}%
\begin{tabular}{c}
{\small{}STICERD}\tabularnewline
{\small{}London School of Economics}\tabularnewline
{\small{}Houghton Street }\tabularnewline
{\small{}London, WC2A 2AE, UK }\tabularnewline
\multicolumn{1}{c}{{\small{}email: f.cowell@lse.ac.uk}}\tabularnewline
\multicolumn{1}{c}{}\tabularnewline
\end{tabular}{\Large\par}
\par\end{center}

\begin{center}
and
\par\end{center}

\begin{center}
{\LARGE{}\medskip{}
}{\LARGE\par}
\par\end{center}

\begin{center}
{\Large{}{}\textbf{\Large{}Emmanuel Flachaire}{\Large{}} }{\Large\par}
\par\end{center}

\begin{center}
{\Large{}}%
\begin{tabular}{c}
Aix-Marseille University \tabularnewline
AMSE \& CNRS\tabularnewline
 5 bd Maurice Bourdet\tabularnewline
{\small{}13001 Marseille, France}\tabularnewline
\multicolumn{1}{c}{{\small{}email: emmanuel.flachaire@univ-amu.fr}}\tabularnewline
\multicolumn{1}{c}{}\tabularnewline
\end{tabular}{\Large\par}
\par\end{center}

{\Large{}\bigskip{}
 }{\Large\par}
\begin{center}
February 2025
\par\end{center}

\vfill{}

\end{titlepage}\pagebreak

\begin{center}
\textbf{Abstract} 
\par\end{center}

\noindent We examine whether mobility measures appropriately represent
changes in individual status, like income or ranks. We suggest three
elementary principles for mobility comparisons and show that many
commonly used indices violate one or more of them. These principles
are used to characterise two classes of measures that have a natural
interpretation in terms of distributional analysis. Class-1 measures
are based on the summation of power functions of individual status
levels and have connections with generalised-entropy and Kolm inequality
measures. Class-2 measures are based on the weighted aggregation of
individual status changes and have connections with (extended) Gini
inequality measures.

\bigskip{}

\bigskip{}
\begin{lyxcode}
\noindent \bigskip{}
\end{lyxcode}
\noindent \textbf{Acknowledgments}: We thank Hyein Cho and Xinyue
Wu for research assistance. Emmanuel Flachaire acknowledges research
support from the French government under the France 2030 investment
plan managed by the French National Research Agency (reference: ANR-17-EURE-0020)
and from Excellence Initiative of Aix-Marseille University - A{*}MIDEX.

\noindent \pagebreak\pagebreak

\section{Introduction}

Evidence of individual mobility is often seen as a desirable objective
for social and economic policy. It is also used indirectly as part
of the discussion of equality of opportunity. Improving data on intra-
and inter-generational mobility has greatly advanced the understanding
of the strengths and limitations of mobility analysis. However, convincing
evidence of mobility requires not only good data but also measurement
tools that have appropriate properties. Perhaps surprisingly, several
commonly-used techniques and indices do not appear to conform well
to simple principles concerning mobility and immobility. 

This paper shows which types of mobility measures are suitable for
the purpose of implementing conventional notions about the meaning
of mobility comparisons. Using the framework introduced in \citeN{CoFl:18MM}
we show that there are two broad classes of mobility measures that
generally satisfy a minimal set of requirements for mobility comparisons.
Each of these classes has a natural interpretation in terms of distributional
analysis and has properties related to those of well-known inequality
indices. 

The paper is organised as follows. Section \ref{sec:concepts-measures}
sets out some general principles on the meaning of mobility comparisons
and examines how well some of the standard tools work in the light
of those principles. Section \ref{sec:theory} provides a theoretical
treatment that embodies the principles of section \ref{sec:concepts-measures}
in a set of axioms and derives a characterisation of two classes of
mobility measures from the axioms. Section \ref{sec:discussion} shows
how these broad characterisations can be embodied in two classes of
measures that can be easily implemented empirically. Section \ref{sec:conclusion}
concludes.

\section{Mobility concepts and measures\label{sec:concepts-measures} }

What do we want a mobility measure to do? Let us discuss this within
the general context of changes in personal status, although the discussion
could also be couched in specific terms of income, wealth, health
or something else. Assume that there are two periods, labelled 0 and
1, and a given number of persons with a known status in periods 0
and 1. ``Status'' could be something very simple, like income, or
something derived from the data on the distribution, such as ordinal
rank in the distribution. For each person we refer to the pair (status-in-0,
status-in-1) as the person's \emph{history}; if we had intergenerational
mobility in mind we might want to refer to the history of a dynasty.

Now let us go through a very short list of principles. 

\subsection{Principles\label{subsec:Principles}}

\paragraph*{1 More movement, more mobility. }

There are two principal interpretations of this concept: (1) more
movement in one person's history (or in one dynasty's history), or
(2) more matched movement-in-pairs history. The reason for considering
two versions of this principle is that each captures a different concept
of mobility. Version (1) allows one to investigate the mobility associated
with unbalanced growth: see, for example the discussion in \citeN{Bour:11NA}.
Version (2) controls for changes in status where the marginal distributions
in periods 0 and 1 have the same mean, so that mobility is associated
with the movement of \emph{shares}; this idea includes both changes
in the marginal distributions that may be reflected in inequality
and the standard interpretations of the concept of ``exchange mobility''
(\citeNP{JaJe:15}; \citeNP{KeGr:81}, page 54; \citeNP{Mccl:77SA})
as pure movement of position. One or other interpretation of this
principle seem to be an almost essential requirement for mobility
measurement. The reason is that in each interpretation this principle
ensures that a mobility measure has a minimum-mobility property: a
situation where there is some movement in status registers higher
mobility than a situation where there is a complete absence of movement.
Of course, the intellectual approach to mobility need not assign the
same importance to every type of movement, as we discuss below.

\paragraph*{2 Decomposition. }

Decomposition analysis is routinely applied to other aspects of distributional
analysis such as income inequality. Several aspects of decomposability
-- such as decomposition by population characteristics -- seem to
be attractive. What has been argued as extremely important in the
context of mobility is the ability to decompose mobility in terms
of upward and downward movements \cite{BaCa:18}. It may also be desirable
to decompose mobility functionally: breaking down overall mobility
into contributions that arise from structural mobility, exchange mobility
and growth. We discuss these issues in section \ref{subsec:Decomposability}. 

\paragraph*{3 Consistency in comparisons. }

It is useful for mobility comparisons to have a consistency property.
A mobility comparison involves comparing one bivariate distribution
of (status-in-0, status-in-1) with another. Suppose one such pair
of distributions is clearly ``similar'' to another -- for example
where the pair of bivariate distributions in the second case can be
found by a simple transformation of the pair of distributions in the
first case, perhaps by just rescaling all the status values by a common
factor, or by just translating the distributions by increasing/decreasing
all the status values by the same given amount.\footnote{Note that this consistency-in-comparisons property does not imply
that the value of a mobility index should be constant under scale
or translation changes.} Then the mobility-ranking for the first pair of distributions should
be the same as for the second pair.\medskip{}

Some approaches to mobility introduce a further principle, based on
the concept of \emph{directionality. }Directionality means that mobility
comparisons should treat differently movements of status in different
directions The question naturally arises, how should directionality
be incorporated in the analysis? Should one introduce the directionality
property explicitly and use special measurement tools for different
mobility problems?\footnote{Examples of this are considered in section \ref{subsec:Directional-mobility-indices}
below. } Should the concept be linked to inequality changes? Should one adopt
an approach that accommodates directionality within a framework set
by other principles of mobility? We use this third, implicit, approach
to directionality: section \ref{subsec:Decomposability} shows how
it emerges naturally from decomposition analysis.

It is useful to examine whether the tools that are conventionally
used to study mobility conform to the three principles monotonicity,
decomposition and consistency: this is discussed in sections \ref{subsec:Statistical-indices}
to \ref{subsec:Checking-the-indices}. 

\subsection{Statistical indices\label{subsec:Statistical-indices}}

Many empirical studies use off-the-shelf tools borrowed from statistics
and applied econometrics. To investigate these let income be denoted
$y$ and assume that status is given by $x=\log\left(y\right)$, so
that the history of person $i$, or dynasty $i$, consists of a log-income
pair $\left(x_{0i},x_{1i}\right)$. There are two standard ``statistical''
indices that are in wide usage. 

One of the most commonly used mobility indices is $1-\hat{\beta}$,
where $\hat{\beta}$ is an\emph{ elasticity coefficient}, computed
as the OLS estimate of the slope coefficient from a linear regression
of log-income in period 1 ($x_{1}$) on log-income on period 0 ($x_{0}$):
\begin{equation}
x_{1i}=\alpha+\beta x_{0i}+\varepsilon_{i}.\label{eq:elast-coeff}
\end{equation}
A high value of $1-\beta$ is usually taken as evidence of significant
mobility. However, a low value does not imply low mobility. It is
easy to construct examples for which $1-\hat{\beta}=0$ but where
common sense suggests that there is indeed mobility in log incomes.
To see this note that, since 
\begin{equation}
\hat{\beta}=\frac{cov(\mathbf{x}_{0},\mathbf{x}_{1})}{var(\mathbf{x}_{0})},\label{eq:beta^}
\end{equation}
$1-\hat{\beta}=0$ if and only if $cov(\mathbf{x}_{0},\mathbf{x}_{1})=var(\mathbf{x}_{0})$.
So, with equidistant log-incomes $\mathbf{x}_{0}=(x_{01},x_{01}+k,x_{01}+2k)$
in period 0, and with log-incomes $\mathbf{x}_{1}=(x_{11},x_{12},x_{11}+2k)$
in period 1, one has $1-\hat{\beta}=0$, $\forall x_{01},x_{11},x_{12}$.\footnote{With equidistant values, mean log-income in period 0 is $\mu_{0}=x_{01}+k$
and one has $x_{01}-\mu_{0}=\mu_{0}-x_{03}=-k$. So $cov(\mathbf{x}_{0},\mathbf{x}_{1})=var(\mathbf{x}_{0})$
is equivalent to $(x_{01}-\mu_{0})[(x_{11}-\mu_{1})-(x_{13}-\mu_{1})]=2(x_{01}-\mu_{0})^{2}$,
which can also be written $x_{11}-x_{13}=-2k$.} For instance, suppose the log-incomes in the two periods 0 and 1
are given by:

\begin{equation}
\begin{cases}
\mathbf{x}_{0}=(1,2,3),\\
\mathbf{x}_{1}\in\{(2,0,4),(2,1,4),(2,1760,4),(2100,1,2102),(2100,74,2102),\dots\};
\end{cases}\label{eq:example_elast_coeff}
\end{equation}
the mobility index based on the elasticity coefficient suggests no
mobility: in all cases  $1-\hat{\beta}=0$. 

Now consider another widely used mobility index, $1-\hat{\rho}$,
where $\hat{\rho}$ is the \emph{Pearson correlation coefficient}
of log-income, 
\begin{equation}
\hat{\rho}=\frac{\sum_{i=1}^{n}\left[x_{0i}-\mu_{0}\right]\left[x_{1i}-\mu_{1}\right]}{\sqrt{\sum_{i=1}^{n}\left[x_{0i}-\mu_{0}\right]^{2}\sum_{i=1}^{n}\left[x_{1i}-\mu_{1}\right]^{2}}},\label{eq:corr-coeff}
\end{equation}
where $\mu_{0},\mu_{1}$ are the means of $\mathbf{x}_{0},\mathbf{x}_{1}$
respectively. This index is both scale independent and translation
independent, that is, if $x_{1}=ax_{0}+b$ then $1-\hat{\rho}=0$.
If $x_{1i}=ax_{0i}+b$ across all $i$ then $\hat{\rho}=1$, so that
mobility ($1-\hat{\rho}$) is zero. For instance, with log-incomes
$\mathbf{x}_{0}=(1,2,3)$ in period 0 and $\mathbf{x}_{1}=(0,2,4)$
in period 1, one has $x_{1}=2x_{0}-2$ and, thus, $1-\hat{\rho}=0$.
This index can also behave strangely. Indeed, with equidistant log-incomes
$\mathbf{x}_{0}=(x_{01},x_{01}+k,x_{01}+2k)$ in period 0, and with
period-1 log-incomes $\mathbf{x}_{1}=(x_{11},x_{12},x_{11})$, one
has $1-\hat{\rho}=1$ and $1-\hat{\beta}=1$, $\forall x_{01},x_{11},x_{12}$.\footnote{With equidistant values, mean log-income in period 0 is $\mu_{0}=x_{01}+k$
and one has $x_{01}-\mu_{0}=\mu_{0}-x_{03}$. Thus, $cov(\mathbf{x}_{0},\mathbf{x}_{1})=k(x_{13}-x_{11})$.
If $x_{13}=x_{11}$then $cov(\mathbf{x}_{0},\mathbf{x}_{1})=0$ ;
therefore $\hat{\beta}=\hat{\rho}=0$} 

So, both statistical measures behave oddly when applied to mobility
problems. For instance, with log-incomes in the two periods given
by 
\begin{equation}
\begin{cases}
\mathbf{x}_{0}=(1,2,3),\\
\mathbf{x}_{1}\in\{(3,2,3),(3,0,3),(3,100,3),(1,2,1),(10,1,10),(2,1,2),(2,100,2),\dots\}
\end{cases},
\end{equation}
the two statistical mobility indices indicate that there is \emph{identical}
mobility: in all cases one has $1-\hat{\rho}=1$ and $1-\hat{\beta}=1$.
Again, take the cases depicted in Table \ref{tab:statistical-mobility-measures}.
It is clear that case 2 exhibits more income movement. Indeed, from
period 0 to 1, log-income variations ($\mathbf{x}_{1}-\mathbf{\mathbf{x}_{0}}$)
are equal to $(2,0,0)$ in case 1 and to $(2,-1,2)$ in case 2. However,
using the mobility indices, $1-\hat{\rho}$ and $1-\hat{\beta}$,
one finds more mobility in case 1. The index based on elasticity even
suggests \emph{no mobility} in case 2.

\begin{table}
\noindent \centering{}%
\begin{tabular}{lcccc}
\hline 
 & period 0  & period 1  & \multicolumn{2}{c}{mobility}\tabularnewline
 & $\mathbf{x}_{0}$  & $\mathbf{x}_{1}$  & $1-\hat{\rho}$  & $1-\hat{\beta}$ \tabularnewline
\hline 
\hline 
case 1  & $(1,2,3)$  & $(3,2,3)$  & 1.0  & 1.0\tabularnewline
case 2  & $(1,2,3)$  & $(3,1,5)$  & 0.5  & 0.0\tabularnewline
\hline 
\end{tabular}\caption{\label{tab:statistical-mobility-measures} statistical mobility indices}
\end{table}

\subsection{Mobility indices connected to inequality measurement\label{subsec:ineq-related-measures}}

\noindent In addition to the well-known statistical mobility indices
discussed in section \ref{subsec:Statistical-indices}, we consider
three mobility indices that have their roots in inequality measurement.
\citeANP{FiOk:96} (\citeyearNP{FiOk:96}, \citeyearNP{FiOk:99MM})
provided indices that are based on differences in income or log-income:
\begin{eqnarray}
FO_{1} & = & \frac{1}{n}\sum_{i=1}^{n}\mid y_{0i}-y_{1i}\mid,\label{eq:FO1}\\
FO_{2} & = & \frac{1}{n}\sum_{i=1}^{n}\mid\log y_{1i}-\log y_{0i}\mid;\label{eq:FO2}
\end{eqnarray}
 $FO_{1}$ and $FO_{2}$ are closely related to the inequality measures
discussed in section \ref{subsec:Mobility-and-inequality}. For these
two indices ``no mobility'' is defined as the situation in which
incomes in both periods are ($FO_{1}$) shifted by the same value
or ($FO_{2}$) multiplied by the same value.

\noindent \citeN{Shor:78II} introduced a family of mobility indices
explicitly based on inequality: 
\begin{equation}
S_{I}=1-\frac{I(y_{0}+y_{1})}{\frac{\mu_{y_{0}}}{\mu_{y_{0}+y_{1}}}\,I(y_{0})+\frac{\mu_{y_{1}}}{\mu_{y_{0}+y_{1}}}\,I(y_{1})},\label{eq:Shorrocks}
\end{equation}
where $I(.)$ is a predefined specified inequality measure. The choice
of inequality measure will crucially affect the behaviour of the mobility
index.

\subsection{Directional mobility indices\label{subsec:Directional-mobility-indices}}

We also consider four recent indices that incorporate the concept
of directionality. \citeN{RaGe:23} proposed an absolute and a relative
index of upward mobility:\footnote{See \citeN{RaGe:23} page 3056 equation (11), and page 3057 equation
(15) respectively.} 
\begin{eqnarray}
RG_{1} & = & -\frac{1}{\alpha}\log\left(\frac{\sum_{i=1}^{n}y_{1i}^{-\alpha}}{\sum_{i=1}^{n}y_{0i}^{-\alpha}}\right),\:\alpha>0,\label{eq:RG1}\\
RG_{2} & = & -\frac{1}{\alpha}\log\left(\frac{\sum_{i=1}^{n}y_{1i}^{-\alpha}}{\sum_{i=1}^{n}y_{0i}^{-\alpha}}\right)+\log\left(\frac{\sum_{i=1}^{n}y_{0i}}{\sum_{i=1}^{n}y_{1i}}\right),\:\alpha>0.\label{eq:RG2}
\end{eqnarray}
$RG_{1}$ and $RG_{2}$ are defined to reward growth and penalise
decline, with higher values when the relatively poor experience faster
growth. In this sense, they are directional and progressive. 

\citeANP{BaCa:18} (\citeyearNP{BaCa:18}, \citeyearNP{BaCa:25})
proposed a downward and an upward mobility index:
\begin{eqnarray}
BC_{\mathrm{D}} & = & \frac{1}{n}\sum_{i\in D}\left(\frac{y_{0i}-y_{1i}}{y_{0i}}\right)^{\alpha},\qquad\alpha\geq0,\label{eq:BC_D}\\
BC_{\mathrm{U}} & = & \frac{1}{n}\sum_{i\in U}\left(\frac{y_{1i}-y_{0i}}{y_{0i}}\right)^{\alpha},\qquad\alpha\geq0,\label{eq:BC_U}
\end{eqnarray}
where $D$ and $U$ are defined as the set of downward and upward
movers, respectively . In constructing $BC_{\mathrm{D}},$ \citeANP{BaCa:25}
effectively censor the bivariate distribution in order to capture
downward mobility. $BC_{\mathrm{D}}$ provides the proportion (incidence,
$\alpha=0$), mean relative income loss (intensity, $\alpha=1$),
and the inequality of relative income losses (inequality, $\alpha=2$)
of downward movers. $BC_{\mathrm{U}}$ is constructed similarly. $BC_{\mathrm{D}}$
and $BC_{\mathrm{U}}$ are both designed to be larger, the larger
is the proportional movement of incomes and the poorer the individuals
affected by it, whenever $\alpha>1$;

\subsection{Checking the indices against the principles\label{subsec:Checking-the-indices}}

Let us summarise the way in which statistical and other mobility indices
do, or do not conform to the principles of mobility measurement set
out in section \ref{subsec:Principles}. Table~\ref{tab:comparisons}
assists with this summary by presenting values of the mobility indices
in different situations involving a three-person world (A, B, C),
in which status is income. Income in period 0 is given as $y_{0}=(10,20,40)$
and there are several scenarios in period 1, with shifted, rescaled
and/or reranked incomes. The Ray-Genicot indices and the B\'arcena-Cant\'o
indices have been computed with the parameter $\alpha$ set to 1.

\begin{table}[htb]
\centering{}{\footnotesize{}{}}%
\begin{tabular}{l@{}c@{}cccccccc}
\hline 
 & period &  & \multicolumn{7}{c}{period}\tabularnewline
\cline{4-10} &  &  &  &  &  &  &  &  & \tabularnewline
 & 0 &  & $1^{a}$ & $1^{b}$ & $1^{c}$ & $1^{d}$ & $1^{e}$ & $1^{f}$ & $1^{g}$\tabularnewline
\hline &  &  &  &  &  &  &  &  & \tabularnewline
A & $10$ &  & $20$ & $15$ & $20$ & $40$ & $25$ & $10$ & $10$\tabularnewline
B & $20$ &  & $40$ & $25$ & $40$ & $80$ & $45$ & $30$ & $40$\tabularnewline
C & $40$ &  & $80$ & $45$ & $10$ & $20$ & $15$ & $40$ & $160$\tabularnewline
\hline &  &  &  &  &  &  &  &  & \tabularnewline
Elasticity & $1-\hat{\beta}$ &  & \textbf{0} & 0.208 & 1.500 & 1.500 & 1.368 & \textbf{0} & -1.000\tabularnewline
Pearson correlation & $1-\hat{\rho}$ &  & \textbf{0} & 0.001 & 1.500 & 1.500 & 1.465 & 0.053 & \textbf{0}\tabularnewline
Fields-Ok 1 & $FO_{1}$ &  & 23.333 & 5.000 & 20.000 & 36.667 & 21.667 & 3.333 & 46.667\tabularnewline
Fields-Ok 2 & $FO_{2}$ &  & 0.693 & 0.249 & 0.924 & 1.155 & 0.903 & 0.135 & 0.693\tabularnewline
Shorrocks 1 & $S_{\text{Theil}}$ &  & \textbf{0} & 0.011 & 0.736 & 0.680 & 0.739 & 0.034 & 0.053\tabularnewline
Shorrocks 2 & $S_{\text{Gini}}$ &  & \textbf{0} & \textbf{0} & 0.500 & 0.444 & 0.500 & \textbf{0} & \textbf{0}\tabularnewline
Ray-Genicot absolute & $RG_{1}$ &  & 0.693 & 0.306 & \textbf{0} & 0.693 & 0.306 & 0.100 & 0.288\tabularnewline
Ray-Genicot relative & $RG_{2}$ &  & \textbf{0} & 0.112 & \textbf{0} & \textbf{0} & 0.112 & -0.033 & -0.811\tabularnewline
B\'arcena-Cant\'o downward & $BC_{\mathrm{D}}$ &  & \textbf{0} & \textbf{0} & 0.250 & 0.167 & 0.208 & \textbf{0} & \textbf{0}\tabularnewline
B\'arcena-Cant\'o upward & $BC_{\mathrm{U}}$ &  & 1.000 & 0.292 & 0.667 & 2.000 & 0.917 & 0.167 & 1.333\tabularnewline
\hline 
\end{tabular}{\footnotesize{}{}\caption{Mobility indices in different scenarios.}
\label{tab:comparisons} }
\end{table}

\emph{More movement, more mobility}. That the statistical indices
violate this principle is clear from the example (\ref{eq:example_elast_coeff})
in section \ref{subsec:Statistical-indices}. Table~\ref{tab:comparisons}
reinforces this point: zero mobility is obtained with scenarios $1^{f}$
or $1^{g}$. It follows that a low value of these indices should not
be used to infer that mobility is low. The Ray-Genicot indices also
violate the movement principle: they are equal to zero in scenario
$1^{c}$, where individual incomes rise or fall and ranks change,
but the set of income values is the same in the two periods:\footnote{The change in the distribution from that in period 0 to that in scenario
$1^{c}$ of period 1 is a simple example of pure exchange mobility,
mentioned above.} these indices ignore a person's history.

\emph{Decomposability}. Inspection of the formulas (\ref{eq:beta^})
to (\ref{eq:BC_U}) shows that the statistical indices, the Shorrocks
and the Ray-Genicot indices are not decomposable; the additive structure
of $FO_{1},$$FO_{2}$, $BC_{\mathrm{D}}$ and $BC_{\mathrm{U}}$
ensure that they are decomposable. 

\emph{Consistency in comparisons}. Because the elasticity and correlation
coefficients are defined on log-income the associated mobility indices
will be scale-independent. This is clear in scenario $1^{a}$ which
shows zero mobility for both indices ($1-\hat{\beta}=1-\hat{\rho}=0$)
and in scenarios $1^{c}$ and $1^{d}$ where for each index shows
the same amount of mobility (1.5) in the two scenarios. The Ray-Genicot
relative index ($RG_{2}$) is also scale independent. By contrast
the Fields-Ok mobility indices are not fully scale-independent, and
so have values different from zero in scenario $1^{a}$ ($FO_{1}=23.333$
and $FO_{2}=0.693$) and they have different values in $1^{c}$ and
$1^{d}$. They are not sensitive to how the absolute variations are
distributed across individuals. For instance, the same index value
is obtained from an increase of only one (log-)income by $n$ and,
from an increase of $n$ (log-)incomes by $1$. In Table~\ref{tab:comparisons},
$FO_{2}$ takes the same value in scenarios $1^{a}$ and $1^{g}$,
which have different movements of income but share the same aggregated
log-income variation. The Shorrocks indices are not scale-independent
(scenarios $1^{c}$ and $1^{d}$ provide different values). Likewise
the \citeANP{BaCa:18} indices are not scale independent in Table
\ref{tab:comparisons}, using $\alpha=1$, $BC_{\mathrm{D}}$ and
$BC_{\mathrm{U}}$ return different values in scenarios $1^{c}$,
$1^{d}$ and $1^{e}$. These indices are sensitive to proportional
income changes throughout the distribution and to translation of incomes.\footnote{However, all the indices except $FO_{1}$ do have the following consistency
property regarding scale changes: if the incomes in both periods were
to be rescaled by the same amount then measured mobility remains unchanged.
This issue is discussed further in section \ref{sec:theory}.} 

Two other features concerning the Shorrocks index should be noted
from Table \ref{tab:comparisons}. First, the index is indeed sensitive
to the choice of inequality measure: Table~\ref{tab:comparisons}
gives very different results with the Theil and Gini indices ($S_{\text{Theil}}$,
$S_{\text{Gini}}$). Second, the Shorrocks index based on the Gini
may appear to be an appropriate index of rank mobility \shortcite{AaBjJaPaPeSmWe:02},
because it is equal to zero when there is no reordering of individual
positions between periods (scenarios $1^{a},1^{b},1^{f}$ and $1^{g}$);
but different values can be returned with the same reranking (scenarios
$1^{c}$ and $1^{d}$). 

\section{Mobility measures:  theory\label{sec:theory}}

In this section we develop the general principles discussed in section
\ref{subsec:Principles} to formalise the principles into axioms and
then using the axioms to characterise mobility orderings. As in section
\ref{sec:concepts-measures} we deal with the problem of two-period
mobility and a fixed population.

\subsection{Status, histories, profiles}

The fundamental ingredient is \emph{status}, which could be a person's
income or wealth, a person's position in the distribution, or something
else: the choice of status concept can be used to pick up different
aspects of the mobility phenomenon and will play a role in the type
of mobility measure that emerges from the analysis. Let $u_{i}$ denote
$i$'s status in period 0 and $v_{i}$ denote $i$'s status in period
1. An individual \emph{history} is the pair $z_{i}=\left(u_{i},v_{i}\right)$.
Individual movements or changes in status are completely characterised
by the histories $z_{i}$, $i=1,2,...,n$. Call the array of such
histories $\mathbf{z}:=\left(z_{1},...,z_{n}\right)$ a \textit{movement
profile},\footnote{The profile concept here is different from that developed in \citeN{VanK:09} }
and denote the set of all possible movement profiles for a population
of size $n$ as $Z^{n}$. The labelling of the histories is arbitrary:
for convenience we adopt, for now, the convention that the labelling
is by the size of the $u$-component so that $u_{1}\leq u_{2}\leq...\leq u_{n}.$ 

The principal problem concerns the representation of the changes from
period 0 to period 1 that are embedded in a given movement profile.
Overall mobility for any member of $Z^{n}$ can be described in terms
of the status changes of each the $n$ persons or dynasties. We need
to specify a set of axioms for comparing the elements of $Z^{n}$
that capture the principles in section \ref{sec:concepts-measures}. 

\subsection{Mobility ordering: basic structure\label{subsec:ordering-structure}}

In this section and section \ref{subsec:ordering-scale} we characterise
an ordering that enables us to compare movement profiles. Use $\succeq$
to denote a weak ordering on $Z^{n}$ that has the standard properties
of completeness and transitivity; denote by $\succ$ the strict relation
associated with $\succeq$ and denote by $\thicksim$ the equivalence
relation associated with $\succeq$. We first consider the interpretation
of the axioms that underpin the approach; we then state a basic result
that follows from them.

\begin{axiom} \textbf{\emph{{[}Continuity{]} }}\label{ax:continuity}$\succeq$
is continuous on $Z^{n}$. \end{axiom}

\begin{figure}
\noindent \centering{}\includegraphics[scale=0.5]{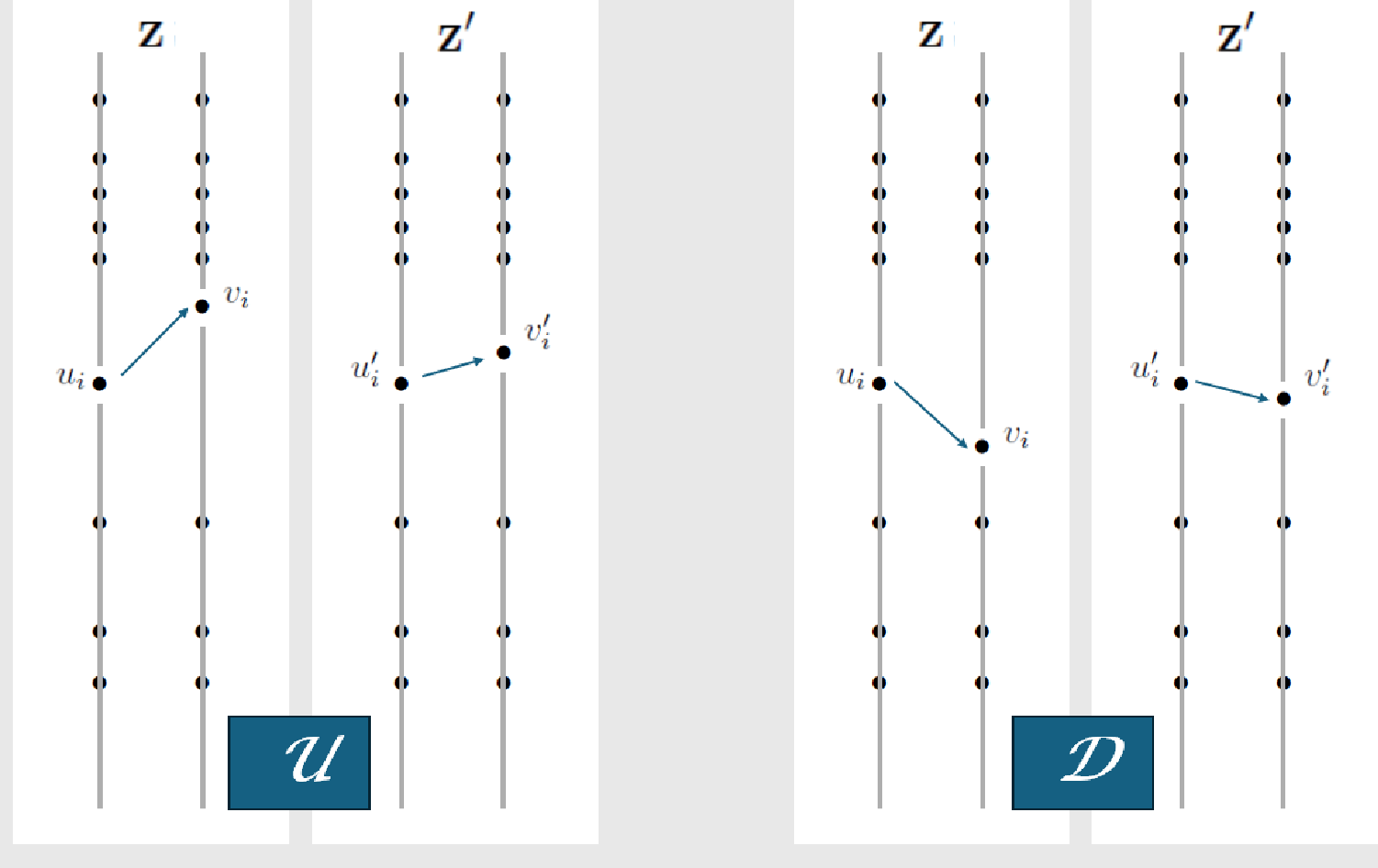}\\
\caption{Monotonicity\label{fig:Monotonicity}}
\end{figure}

\begin{axiom} \textbf{\emph{{[}Monotonicity{]} }}\label{ax:monotonicity}Let
$\mathbf{z,z}^{\prime}\in Z^{n}$ differ only in their $i$th history
and $u_{i}^{\prime}=u_{i}$ and define two conditions $\mathcal{U}:=\lyxmathsym{\textquotedblleft}v_{i}>v_{i}^{\prime}\geq u_{i}\lyxmathsym{\textquotedblright}$
and $\mathcal{D}:=\text{\textquotedblleft}v_{i}<v_{i}^{\prime}\leq u_{i}"$.
If $z_{i}$ satisfies either $\mathcal{U}$ or $\mathcal{D}$ then 

$\mathbf{z}\succ\mathbf{z}^{\prime}$. \end{axiom} The interpretation
of Axiom \ref{ax:monotonicity} is as follows. Suppose that, with
the sole exception of person $i$, each person's history in profile
$\mathbf{z}$ is the same as it is in profile $\mathbf{z}^{\prime}.$
Person $i$'s history can be described thus: $i$ starts with the
same period-0 status in $\mathbf{z}$ and in $\mathbf{z}^{\prime}$
and then moves up to a higher status in period 1; but $i$'s period-1
status in profile $\mathbf{z}$ is even higher than it is in $\mathbf{z}^{\prime}$.
Then Axiom \ref{ax:monotonicity} implies that mobility would be higher
in $\mathbf{z}$ than in $\mathbf{z}^{\prime}$; a corresponding story
holds for downward movement: more movement implies more mobility.
The two conditions $\mathcal{U}$ and $\mathcal{D}$ are illustrated,
respectively, in the left-hand and right-hand panels of Figure \ref{fig:Monotonicity},
where each pair of columns represents the status values in the two
periods 0 and 1. It should also be noted that two modified versions
of Axiom \ref{ax:monotonicity} may also be of interest: one version
would use the wording ``if $z_{i}$ satisfies $\mathcal{U}$ (alone)''
in order to focus solely on upward mobility; the other version would
use the wording ``if $z_{i}$ satisfies $\mathcal{D}$ (alone)''
in order to focus solely on downward mobility.  

\begin{figure}
\noindent \centering{}\includegraphics[scale=0.5]{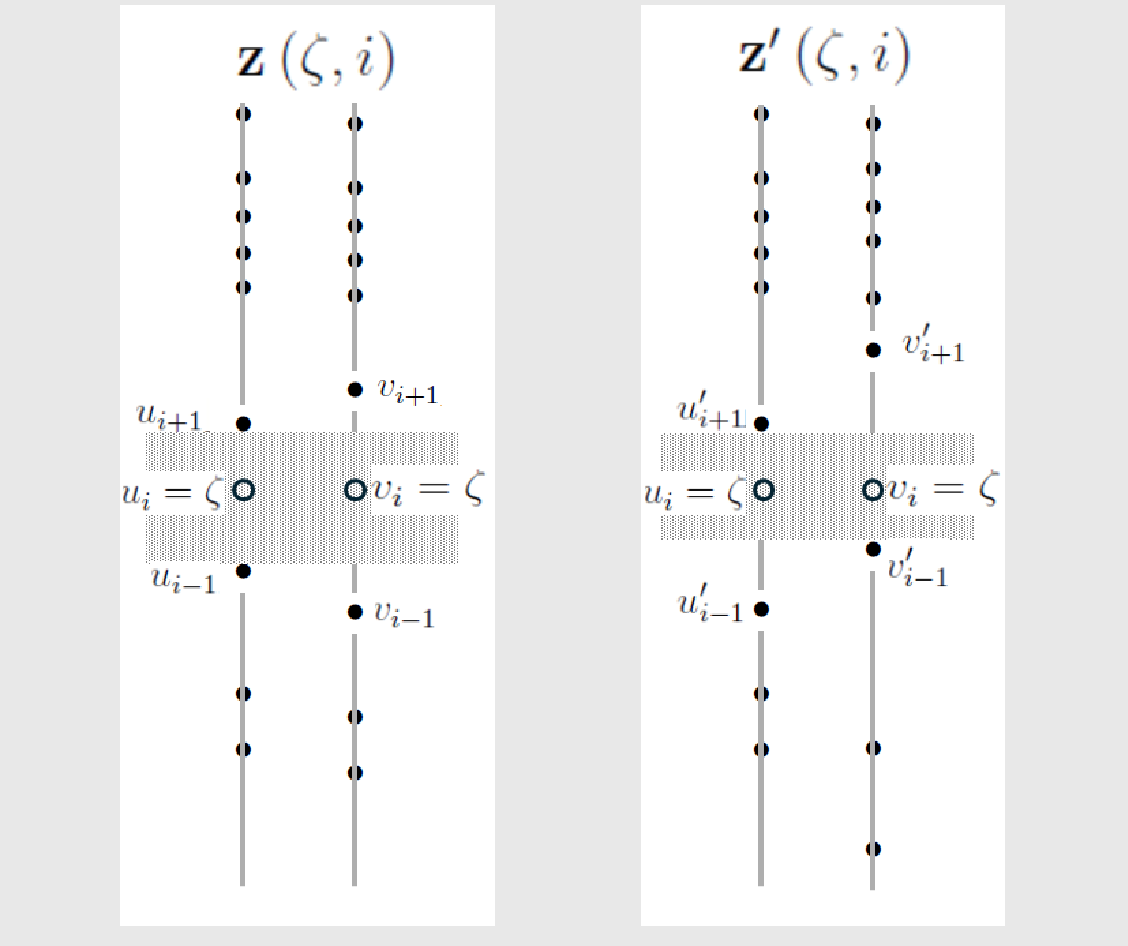}\\
\caption{Independence\label{fig:Independence}}
\end{figure}

\begin{axiom} \textbf{\emph{{[}Independence{]}}} \label{ax:independence}Consider
two profiles $\mathbf{z,z}^{\prime}\in Z^{n}$ where there is some
$i\in\left\{ 2,...,n-1\right\} $ such that $u_{i-1}<u_{i}<u_{i+1}$,
$v_{i-1}<v_{i}<v_{i+1}$, $u'_{i-1}<u_{i}<u'_{i+1}$, $v'_{i-1}<v_{i}<v'_{i+1}$.
Let $\mathbf{z}\left(\boldsymbol{\zeta},i\right)$ denote the profile
formed by replacing the $i$th history in $\mathbf{z}$ by the history
$\boldsymbol{\zeta}\in Z$ and let $\hat{Z}_{i}:=\left[u_{i-1},u_{i+1}\right]\times\left[v_{i-1},v_{i+1}\right]$.
If $\mathbf{z}\thicksim\mathbf{z}^{\prime}$ and $z_{i}=z_{i}^{\prime}$
then $\mathbf{z}\left(\boldsymbol{\zeta},i\right)\thicksim\mathbf{z}^{\prime}\left(\boldsymbol{\zeta},i\right)$
for all $\boldsymbol{\zeta}\in\hat{Z}_{i}\bigcap\hat{Z}'_{i}$ .  \end{axiom}
To interpret this, suppose that the profiles $\mathbf{z}$ and $\mathbf{z}^{\prime}$
are considered to be equivalent in terms of overall mobility and that
there is some person $i$ who has the same history $z_{i}=\left(u_{i},v_{i}\right)$
in both $\mathbf{z}$ and $\mathbf{z}^{\prime}$ and whose status
is strictly greater than that of $i-1$ and strictly less than that
of $i+1$, in both periods. Consider a change in $i$'s history small
enough that these strict inequalities still hold in both $\mathbf{z}$
and $\mathbf{z}^{\prime}$; this change would leave unaltered the
person's ranking in both periods. Axiom \ref{ax:independence} requires
that such a change leaves the two modified profiles as equivalent
in terms of overall mobility. Figure \ref{fig:Independence} illustrates
this in a convenient special case where $u_{i}=\zeta$ and $v_{i}=\zeta$;
profiles $\mathbf{z}$ and $\mathbf{z}^{\prime}$ are assumed to be
equivalent and, if $\zeta$ is varied within the limits indicated
by the shaded area\footnote{These limits ensure that the variation in $\zeta$ will not result
in any reranking.} then the new profiles $\mathbf{z}\left(\boldsymbol{\zeta},i\right)\text{ and }\mathbf{z}^{\prime}\left(\boldsymbol{\zeta},i\right)$
are also regarded as equivalent in terms of mobility.
 The independence property is of great importance for the decomposition
of mobility measures, allowing one to examine in detail the relative
contributions of upward and downward mobility to overall mobility
-- see section \ref{subsec:Decomposability}. 

\begin{axiom} \textbf{\emph{{[}Local immobility{]}}} \label{ax:local-immobility}Let
$\mathbf{z,z}^{\prime}\in Z^{n}$ where, for some $i$, $u_{i}=v_{i}$,
$u_{i}^{\prime}=v_{i}^{\prime}$ and, for all $j\neq i$, $u_{j}^{\prime}=u_{j}$,
$v_{j}^{\prime}=v_{j}$. Then $\mathbf{z\thicksim z}^{\prime}$. \end{axiom}
Consider a profile $\mathbf{z}$ in which person $i$ is immobile:
if we were to replace $i$'s in both periods with a higher value then
the new profile $\mathbf{z}^{\prime}$ would exhibit the same mobility
as the original $\mathbf{z}$. 

These axioms can then be used to establish:\footnote{The proof this and other results is in the Appendix.}

\begin{theorem} \label{th:additive-structure}Given Axioms \ref{ax:continuity}
to \ref{ax:local-immobility} then $\forall\mathbf{z}\in Z^{n}$ the
mobility ordering $\succeq$\ is representable by an increasing monotonic
transform of 
\begin{equation}
\sum_{i=1}^{n}\phi_{i}\left(z_{i}\right),\label{eq:basic-linear}
\end{equation}
where the $\phi_{i}$ are continuous functions $Z\rightarrow\mathbb{R}$,
defined up to an affine transformation, each of which is increasing
(decreasing) in $v_{i}$ if $v_{i}>(<)\:u_{i}$ and that has the property
$\phi_{i}\left(u,u\right)=b_{i}u,$ where $b_{i}\in\mathbb{R}$.\end{theorem} 

It makes sense to choose $b_{i}=0$, so that
\begin{equation}
\phi_{i}\left(u,u\right)=0.\label{eq:phi_i(u,u)}
\end{equation}
This normalisation condition will ensure that when there is no change
in anyone's history ($u_{i}=v_{i}$ for every $i$) mobility is zero.

\bigskip{}

Theorem \ref{th:additive-structure} shows that mobility comparisons
should be based on the sum of evaluations of the $n$ individual histories;
but these evaluations may differ from person to person and may, accordingly,
depend on the rank-order of the person's history in the movement profile.
Axiom \ref{ax:continuity} ensures the continuity of each of the functions
$\phi_{i}\left(\cdot\right)$. Axiom \ref{ax:monotonicity} ensures
the monotonicity of each function $\phi_{i}\left(\cdot\right)$.\footnote{If either of the alternative versions of Axiom \ref{ax:monotonicity}
is adopted then this statement must be modified to allow for cases
where changes in $v_{i}$ leave the assessment of mobility unchanged.} Axiom \ref{ax:independence} yields the additive structure of equation
(\ref{eq:basic-linear}) and allows some form of decomposability of
mobility measures.

\subsection{Mobility ordering: consistency\label{subsec:ordering-scale}}

The second part of our characterisation of the mobility ordering involves
the comparison of movement profiles at different levels of status.
To do this we examine profiles that are regarded as equally mobile,
in other words profiles $\mathbf{z,z}^{\prime}\in Z^{n}$ such that
the following holds:

\begin{equation}
\left.\begin{array}{c}
\mathbf{z\thicksim z}^{\prime}\\
\left(\mathbf{u,v}\right)\sim\left(\mathbf{u^{\prime},v^{\prime}}\right)
\end{array}\right\} .\label{eq:equivalent_profiles}
\end{equation}

\subsubsection{Scale consistency\label{subsec:Consistency-concepts}}

It is useful to distinguish the following three alternative concepts
that could be reasonably invoked to characterise consistency when
status is rescaled. In each case (\ref{eq:equivalent_profiles}) is
true and $\lambda$ is an arbitrary positive number, 

\begin{equation}
\text{\textbf{Origin\;scale\;invariance\;(OSI):}}\left(\lambda\mathbf{u,v}\right)\sim\left(\lambda\mathbf{u^{\prime},v^{\prime}}\right).\label{eq:OSI}
\end{equation}
\begin{equation}
\text{\textbf{Destination\ scale\ invariance\ (DSI):}}\left(\mathbf{u,\lambda v}\right)\sim\left(\mathbf{u^{\prime},\lambda v^{\prime}}\right)\label{eq:DSI}
\end{equation}
\begin{equation}
\textbf{Profile\ scale\ invariance\ (PSI)}:\mathbf{\lambda z}\thicksim\lambda\mathbf{z}^{\prime}\label{eq:PSI}
\end{equation}
The interpretation of OSI and DSI is that the ordering of profiles
by mobility remains unchanged by a scale change in status in one or
other of the two periods. Likewise, PSI means that the ordering of
profiles remains unchanged by a simultaneous scale change to status
in both periods. We have the following results for these three concepts:\footnote{The proofs of the theorems are in the Appendix.}

\begin{theorem}\label{th:Relation-between-axioms} If any two of
the three concepts OSI, DSI, PSI  hold then the third must hold as
well.\end{theorem} 

\begin{theorem}\label{th:scale-three-defs}Given Axioms \ref{ax:continuity}
to \ref{ax:local-immobility} $\succeq$\ is representable by (\ref{eq:basic-linear}),
where $\phi_{i}$ is given by
\begin{equation}
\phi{}_{i}\left(u,v\right)=\begin{cases}
A_{i}\left(v\right)\left[u^{\alpha}-v^{\alpha}\right] & \text{if OSI holds,}\\
A'_{i}\left(u\right)\left[v^{\alpha}-u^{\alpha}\right] & \text{if DSI holds},\\
v^{\beta}h_{i}\left(\frac{u}{v}\right) & \text{if PSI holds},
\end{cases}\label{eq:phi_cases}
\end{equation}
where $A_{i},A'_{i},h$ are functions of one variable and $\alpha,\beta$
are constants. \end{theorem} 

\subsubsection{Combining consistency concepts: OSI, DSI and PSI}

There is an important corollary of Theorems \ref{th:Relation-between-axioms}
and \ref{th:scale-three-defs}: requiring that any \emph{two} of the
three scale-consistency concepts (OSI, DSI, PSI) imposes more structure
on the function used to represent the mobility ordering $\succeq$\ .%

From (\ref{eq:phi_cases}) OSI and PSI imply %

\begin{equation}
h_{i}\left(\frac{u}{v}\right)=A_{i}\left(v\right)\left[u^{\alpha}v^{-\beta}-v^{\alpha-\beta}\right].\label{eq:OSI=000026PSI}
\end{equation}

For this to hold the right hand side of (\ref{eq:OSI=000026PSI})
must be a function of $u/v$ which, for a non-trivial solution, means
that $A_{i}\left(v\right)=c_{i}v^{\beta-\alpha}$ so that
\begin{equation}
\phi{}_{i}\left(u,v\right)=c_{i}v^{\beta}\left[\left[\frac{u}{v}\right]^{\alpha}-1\right],\label{eq:evaluation_function_PSI+OSI}
\end{equation}
where $c_{i}$ is a constant specific to each history $i$. 

\subsubsection{Translation consistency\label{subsec:Translation-consistency}}

We can also consider an alternative consistency concept that focuses
on what happens to orderings if distributions are shifted by a given
absolute amount of status rather than proportional rescaling considered
in section \ref{subsec:Consistency-concepts}. Consider three different
concepts relating to the idea of ``translation invariance.'' In
each case (\ref{eq:equivalent_profiles}) is true, $\mathbf{1}$ is
an array of ones, and $\delta$ is any real number: 

\begin{equation}
\text{\textbf{Origin\;translation\;invariance\;(OTI):}}\left(\mathbf{u+\delta1,v}\right)\sim\left(\mathbf{u^{\prime}+\delta1,v^{\prime}}\right).\label{eq:OTI}
\end{equation}
\begin{equation}
\text{\textbf{Destination\ translation\ invariance\ (DTI):}}\left(\mathbf{u,v}+\delta\boldsymbol{1}\right)\sim\left(\mathbf{u^{\prime},v^{\prime}+\delta1}\right).\label{eq:DTI}
\end{equation}
\begin{equation}
\textbf{Profile\ translation\ invariance\ (PTI)}:\mathbf{z+\delta1}\thicksim\mathbf{z}^{\prime}+\delta\boldsymbol{1.}\label{eq:PTI}
\end{equation}
The interpretation of OTI and DTI is that the ordering of profiles
by mobility remains unchanged by a change in status level in one or
other of the two periods. Likewise, PSI means that the ordering of
profiles remains unchanged by a simultaneous level change to status
in both periods. It is again true the if any two of (OTI, DTI, PTI)
hold, the third property also holds. We have the following result
for translation consistency: %

\begin{theorem} \label{th:PTI}Given Axioms \ref{ax:continuity}
to \ref{ax:local-immobility} , the ordering $\succeq$\ is representable
by (\ref{eq:basic-linear}), where $\phi_{i}$ is given by 

\begin{equation}
\phi{}_{i}\left(u,v\right)=\begin{cases}
A_{i}\left(v\right)\frac{1}{\beta}\left[e^{\beta\left[u-v\right]}-1\right] & \text{if OTI holds,}\\
A'_{i}\left(u\right)\frac{1}{\beta}\left[e^{\beta\left[v-u\right]}-1\right] & \text{if DTI holds},\\
\text{either\ }e^{au}g_{i}\left(v-u\right)\ \text{or\ }e^{av}g'_{i}\left(u-v\right) & \text{if PTI holds},
\end{cases}\label{eq:phi_PTI}
\end{equation}
where $A_{i},A'_{i},g_{i},$$g'_{i}$ are functions of one variable,
$\beta$ and $a$ are constants.\end{theorem} 

Note that the form $\frac{1}{\beta}\left[e^{\beta\left[v-u\right]}-1\right]$
becomes $v-u$ as $\beta\rightarrow0$ -- see footnote \ref{fn:The-normalisation-is}
in the Appendix. The function $g_{i}$ in (\ref{eq:phi_PTI}) is arbitrary
and again it is useful to consider imposing an additional consistency
requirement to obtain more structure. This can be done in two ways.

First, if OTI and DTI hold jointly (which also implies that PTI holds)
the solutions in (\ref{eq:phi_PTI}) imply 
\begin{equation}
\phi{}_{i}\left(u,v\right)=A_{i}\left(v\right)\frac{1}{\beta}\left[e^{\beta\left[u-v\right]}-1\right]=A'_{i}\left(u\right)\frac{1}{\beta}\left[e^{\beta\left[v-u\right]}-1\right],\label{eq:phi_OTI=000026DTI}
\end{equation}
which implies $A'_{i}\left(u\right)=-A_{i}\left(v\right)e^{\beta\left[u-v\right]}$.
If DTI and PTI hold (\ref{eq:phi_PTI}) also implies: 

\begin{equation}
g_{i}\left(v-u\right)=e^{-au}A'_{i}\left(u\right)\frac{1}{\beta}\left[e^{\beta\left[v-u\right]}-1\right].\label{eq:DTI=000026PTI}
\end{equation}
If (\ref{eq:DTI=000026PTI}) is to hold for arbitrary $u,v$ then
$e^{-au}A'_{i}\left(u\right)=a_{i}$, a constant, so that $A'_{i}\left(u\right)=e^{au}a_{i}$.
Therefore, for all $\beta$: 
\begin{equation}
\phi{}_{i}\left(u,v\right)=e^{au}g_{i}\left(v-u\right)=e^{au}a_{i}\frac{1}{\beta}\left[e^{\beta\left[v-u\right]}-1\right].\label{eq:phi_DTI=000026PTI}
\end{equation}
Likewise we have, for all $\beta$:
\begin{equation}
\phi{}_{i}\left(u,v\right)=e^{av}g'_{i}\left(u-v\right)=e^{av}a'_{i}\frac{1}{\beta}\left[e^{\beta\left[u-v\right]}-1\right].\label{eq:phi_DTI=000026PTI'}
\end{equation}
In the case where $\beta=0$ equations (\ref{eq:phi_DTI=000026PTI})
and (\ref{eq:phi_DTI=000026PTI'}) imply
\begin{equation}
\phi{}_{i}\left(u,v\right)=e^{au}a_{i}\left[v-u\right]=e^{av}a'_{i}\left[u-v\right],\label{eq:phi_DTI=000026PTI_beta=00003D0}
\end{equation}
which further implies $e^{au}a_{i}=-e^{av}a'_{i}$. If this is to
hold for arbitrary $u,v$ then $a=0$ and $a_{i}=-a'_{i}.$ So, for
DTI and PTI to hold jointly, one of the following two conditions (\ref{eq:DTI=000026PTI-1},\ref{eq:DTI=000026PTI-0})
must hold: 
\begin{equation}
\phi_{i}\left(u,v\right)=a_{i}\frac{1}{\beta}\left[e^{\beta\left[v-u\right]}-1\right],\text{with }\text{\ensuremath{\beta\neq0},}\label{eq:DTI=000026PTI-1}
\end{equation}
\begin{equation}
\phi_{i}\left(u,v\right)=a_{i}[v-u]\text{.}\label{eq:DTI=000026PTI-0}
\end{equation}

Second, for mobility orderings, it can make sense to require \emph{both}
scale invariance \emph{and} translation invariance of movement profiles.
Taking the conditions for both PSI and PTI together, Theorems \ref{th:scale-three-defs}
and \ref{th:PTI} would require:

\begin{equation}
v^{\beta}h{}_{i}\left(\frac{u}{v}\right)=e^{au}g{}_{i}\left(v-u\right).\label{eq:PSI+PTI}
\end{equation}
A solution to (\ref{eq:PSI+PTI}) for arbitrary $u$ and $v$ is as
follows: $\beta=1$, $a=0$, $h_{i}\left(x\right)=$$a_{i}\left[1-x\right]$
and $g_{i}\left(x\right)=a_{i}x$. This solution implies that equation
(\ref{eq:DTI=000026PTI-0}) holds.

\bigskip{}
The corollaries to Theorems \ref{th:scale-three-defs} and \ref{th:PTI}
show that the evaluation function $\phi_{i}$ in (\ref{eq:basic-linear})
can take one of the two convenient forms (\ref{eq:evaluation_function_PSI+OSI})
and (\ref{eq:DTI=000026PTI-0}). The solution (\ref{eq:DTI=000026PTI-0})
will be considered further in section \ref{subsec:Class-2}.

\subsection{Aggregate mobility measure\label{subsec:aggregate-mobility}}

So it appears that the mobility ordering $\succeq$ implied by the
axioms in sections \ref{subsec:ordering-structure} and \ref{subsec:ordering-scale}
can be represented by the expression $\sum_{i=1}^{n}\phi_{i}\left(u,v\right),$
with the $\phi{}_{i}$ given by equation (\ref{eq:evaluation_function_PSI+OSI})
or (\ref{eq:DTI=000026PTI-0}), depending on the consistency principles
that are to be applied. Since $\succeq$ is an ordering it is also
representable by some continuous increasing transformation of this
expression. 

We now examine what normalisation is appropriate in order to construct
an aggregate mobility measure, for each of the  cases in Theorems
\ref{th:scale-three-defs} and \ref{th:PTI}. In this section and
section \ref{sec:discussion}. We do this for both class-1 and class-2
measures and we distinguish between absolute ($\mathcal{A}$), scale-independent
($\mathcal{S}$), and translation-independent ($\mathcal{T}$) versions
for each of the two classes.

\subsubsection{Class 1: $\phi{}_{i}$ given by equation (\ref{eq:evaluation_function_PSI+OSI})}

First, let us require that mobility should be blind as to individual
identity. So far we adopted, for convenience, the convention that
the labels $i=1,...,n$ were assigned in ascending order of $u_{i}$.
In aggregating the individual $\phi_{i}\left(u_{i},v_{i}\right)$,
with the function $\phi{}_{i}$ given by (\ref{eq:evaluation_function_PSI+OSI}),
one might want to consider alternative labelling, reflecting perhaps
a special merit attached to some of the $n$ persons. However, if
the definition of status incorporates all relevant information about
an person, the labelling $i=1,...,n$ is irrelevant and anonymity
is an innocuous assumption. This means that mobility depends only
on individual status histories; switching the personal labels from
one history to another within a movement profile has no effect on
mobility rankings: if a profile $\mathbf{z}{}^{\prime}$ can be obtained
as a permutation of the components of another profile $\mathbf{z}$,
then they should be treated as equally mobile. If so, then all the
$c_{i}$ should be equal and, using equation (\ref{eq:evaluation_function_PSI+OSI}).
mobility can be represented as a transform of 

\begin{equation}
c\sum_{i=1}^{n}\left[u_{i}^{\alpha}v_{i}^{\beta-\alpha}-v_{i}^{\beta}\right].\label{eq:Class1_norm}
\end{equation}

Now consider the effect of population size $n$. A simple replication
of profiles $\mathbf{z}$ does not change the essential facts of mobility.
However, the constant $c$ in (\ref{eq:Class1_norm}) may depend on
$n$. If any profile is replicated $r$ times and the measure remains
unchanged under replication equation (\ref{eq:Class1_norm}) yields:
\begin{equation}
c\left(n\right)\sum_{i=1}^{n}\left[u_{i}^{\alpha}v_{i}^{\beta-\alpha}-v_{i}^{\beta}\right]=c\left(nr\right)r\sum_{i=1}^{n}\left[u_{i}^{\alpha}v_{i}^{\beta-\alpha}-v_{i}^{\beta}\right].\label{eq:Class1_norm_2}
\end{equation}
So, to ensure that the representation of $\succeq$ is in a form that
is constant under replication, we need to have $c\left(nr\right)r=c\left(n\right)$
which means that $c\left(n\right)=c'/n$ where $c'$ is a constant. 

Further normalising by setting $\beta=1$ and $c'=\frac{1}{\alpha\left[\alpha-1\right]}$,
we may write the measure implied by equation (\ref{eq:Class1_norm_2})
as some transform of the following ``basic-form'' Class-1 mobility
measure: 
\begin{equation}
\frac{1}{\alpha\left[\alpha-1\right]}\left[\frac{1}{n}\sum_{i=1}^{n}u_{i}^{\alpha}v_{i}^{1-\alpha}-\mu_{v}\right],\label{eq:basic-form-uv}
\end{equation}
where $\mu_{v}:=\frac{1}{n}\sum_{i=1}^{n}v_{i}$ and $\alpha\neq0,1$.

Equation (\ref{eq:basic-form-uv}) can be rewritten as 

\[
\frac{1}{\alpha-1}\left[\frac{1}{n}\sum_{i=1}^{n}v_{i}\left[\frac{\left[u_{i}/v_{i}\right]^{\alpha}-1}{\alpha}\right]\right],
\]
and so, using l'H\^opital's rule, we obtain the limiting form for (\ref{eq:basic-form-uv})
as $\alpha\rightarrow0$: $\frac{1}{n}\sum_{i=1}^{n}v_{i}\log\frac{u_{i}}{v_{i}}.$
However, the limiting form for (\ref{eq:basic-form-uv}) as $\alpha\rightarrow1$
is more difficult. Equation (\ref{eq:basic-form-uv}) can be rewritten
as 
\[
\frac{1}{\alpha\left[\alpha-1\right]}\left[\frac{1}{n}\sum_{i=1}^{n}u_{i}^{\alpha}v_{i}^{1-\alpha}-\mu_{u}\right]+\frac{1}{\alpha\left[\alpha-1\right]}\left[\mu_{u}-\mu_{v}\right],
\]
where $\mu_{u}:=\frac{1}{n}\sum_{i=1}^{n}u_{i}$. The first term converges
to $\frac{1}{n}\sum_{i=1}^{n}u_{i}\log\left(\frac{u_{i}}{v_{i}}\right),$
but the second term does not converge so, unless $\mu_{u}=\mu_{v}$,
(\ref{eq:basic-form-uv}) does not converge as $\alpha\rightarrow1$. 

Drawing these results together we have the following for class-1 \uline{A}bsolute
mobility measures, defined as follows: 
\begin{equation}
\mathcal{A}_{\alpha}^{1}:=\begin{cases}
\frac{1}{\alpha\left[\alpha-1\right]n}\sum_{i=1}^{n}\left[u_{i}^{\alpha}v_{i}^{1-\alpha}-\mu_{v}\right],\  & \alpha\neq0,1,\\
\\
-\frac{1}{n}\sum_{i=1}^{n}v_{i}\log\left(u_{i}/v_{i}\right), & \alpha=0,\\
\\
\frac{1}{n}\sum_{i=1}^{n}u_{i}\log\left(u_{i}/v_{i}\right) & \alpha=1\text{ and \ensuremath{\mu_{u}}=\ensuremath{\mu_{v}},}\\
\\
\infty & \alpha=1\text{ and \ensuremath{\mu_{u}\neq\mu_{v}}}.
\end{cases}\label{eq:mobility-c1-abs}
\end{equation}
The basic form (\ref{eq:mobility-c1-abs}) has the property that mobility
is zero if $v_{i}=u_{i}$ for all $i$. In general, if the normalised
mobility measure in (\ref{eq:basic-form-uv}) is to have this ``zero-mobility''
property, it must take the form 
\begin{equation}
\psi\left(\frac{1}{n}\sum_{i=1}^{n}u_{i}^{\alpha}v_{i}^{1-\alpha}-\theta\left(\mu_{u},\mu_{v}\right),\mu_{u},\mu_{v}\right)\label{eq:basic-form-uv-0}
\end{equation}
where $\psi$ is monotonic in its first argument with the property
that $\psi\left(0,\mu_{u},\mu_{v}\right)=0$, and $\theta$ is homogeneous
of degree 1 with the property that $\theta\left(\mu,\mu\right)=\mu$. 

\subsubsection{Class 2: $\phi{}_{i}$ given by equation (\ref{eq:DTI=000026PTI-0})\label{subsec:Class-2}}

Again consider the issue of anonymity. Equation (\ref{eq:DTI=000026PTI-0})
means that individual mobility for each person $i$ is captured simply
by $a_{i}d_{i}$, where $d_{i}$ is the difference in status between
periods 0 and 1: 
\begin{equation}
d_{i}=v_{i}-u_{i}.\label{eq:simple-difference}
\end{equation}
The overall mobility measure will preserve anonymity if it is written
as $\sum_{i=1}^{n}a_{i}d_{(i)}$ where $d_{(i)}$ denotes the $i$th
component of the vector $\left(d_{1},...,d_{n}\right)$ when rearranged
in ascending order: so, except in the case where there is zero individual
mobility, $d_{(1)}<0$ refers to the greatest downward mobility and
$d_{(n)}>0$ to the greatest upward mobility. The principle of monotonicity
is preserved if $a_{i}<0$ whenever $d_{(i)}<0$ and $a_{i}>0$ whenever
$d_{(i)}>0$. The independence of population size means that the term
$a_{i}$ should be normalised by $1/n$; so, up to a change in scale,
we have the mobility measure
\begin{equation}
\frac{1}{n}\sum_{i=1}^{n}a_{i}d_{(i)}.\label{eq:Mobility-linear}
\end{equation}

\subsubsection{Data issues }

The analysis presented so far is suitable for situations where status
can be represented by cardinal data, such as income and wealth. Here
we consider some of the ways in which the implementation of mobility
measures may be affected by the type of data available.

First, it is important to note that, if the data contain negative
or zero values, some measures belonging to Class 1 will be not be
defined for particular parameter values -- for example cases where
the sensitivity parameter is less than zero in (\ref{eq:basic-form-uv-0}).
All of the Class-2 measures are defined for all status values. 

Second, the analysis can be extended to situations where the original
data are ordinal rather than cardinal -- for example self-reported
health or happiness which are typically reported categorically. The
status concept in such cases can be handled as with measuring inequality
using ordinal data -- the individual's status measure then becomes
the person's position in the distribution \cite{CoFl:17IW}. Of course,
position in the distribution may be of interest as a status measure
even when the underlying data are cardinal. Given that the resulting
status will be strictly positive for each individual, all the Class-1
and Class-2 measures will be defined for this way of handling ordinal
data. 

Third, it is important to consider cases where the data are cardinal
but are subject to the restriction that the mean value is fixed. The
individual observations are shares in a fixed total. Interpretation
1 of the ``movement'' principle in section \ref{sec:concepts-measures}
is not appropriate for such cases. We now address this issue.

\subsection{Mean-Normalisation\label{subsec:Normalisation}}

The mobility measures derived in section \ref{subsec:aggregate-mobility}
are consistent with the first version of the ``More movement, more
mobility'' principle discussed in section \ref{sec:concepts-measures}.
We now consider a modification that will enable us to handle version
(2) of that principle.

This can be done by replacing Axiom \ref{ax:monotonicity} with the
following Axiom 2$'$:\\
\\
\textbf{Axiom $\mathbf{2'}$ {[}Monotonicity-2{]}}\textbf{\emph{ }}\label{ax:monotonicity-2-1}\emph{If
$\mathbf{z,z}^{\prime}\in Z^{n}$ differ only in their $i$th and
$j$th components and $u_{i}^{\prime}=u_{i}$, $u_{j}^{\prime}=u_{j}$,
$v_{i}^{\prime}-v_{i}=v_{j}-v_{j}^{\prime}$ then, if $v_{i}>v_{i}^{\prime}\geq u_{i}$
and if $v_{j}<v_{j}^{\prime}\leq u_{j}$, $\mathbf{z}\succ\mathbf{z}^{\prime}$.}
\begin{figure}
\noindent \centering{}\includegraphics[scale=0.5]{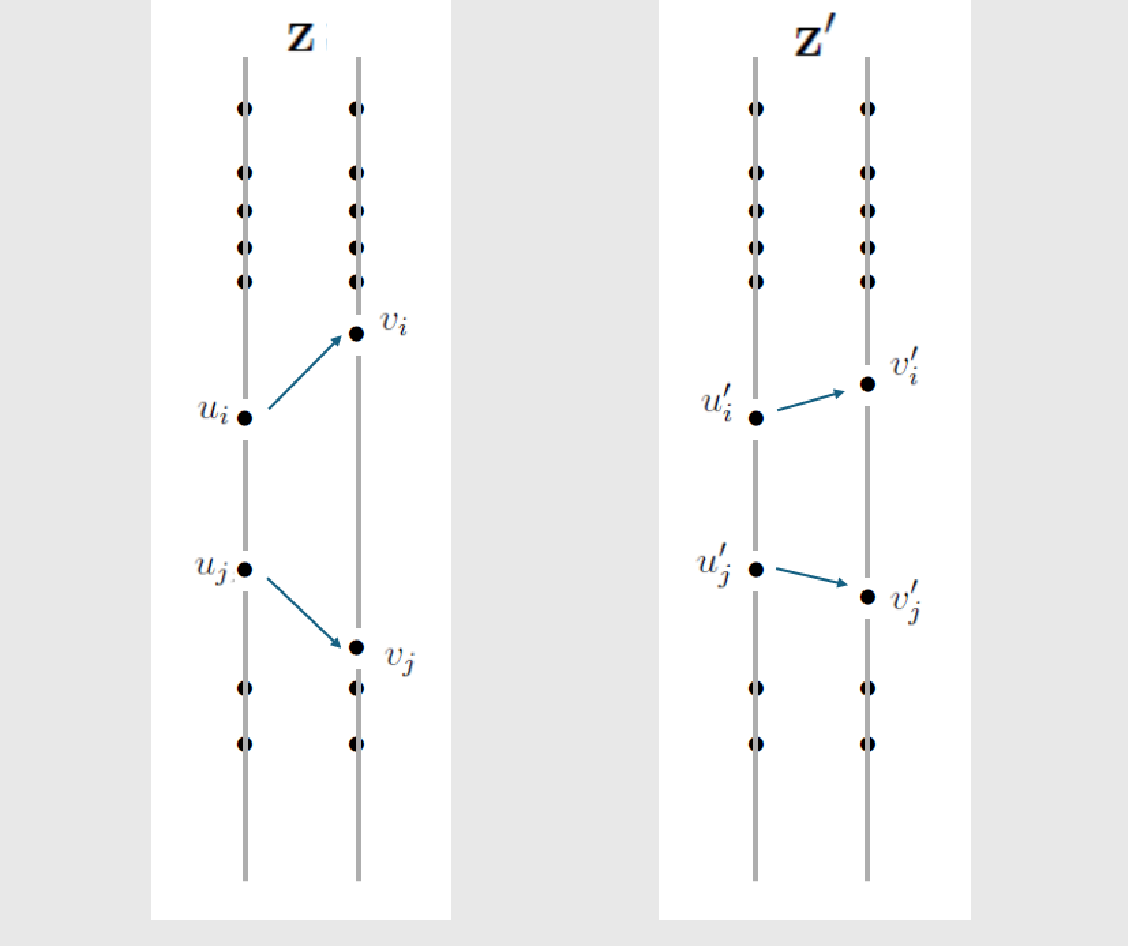}\\
\caption{Monotonicity-2\label{fig:Monotonicity-2}}
\end{figure}

\smallskip{}

The idea behind Axiom 2$'$ is illustrated by Figure \ref{fig:Monotonicity-2},
a modification of Figure \ref{fig:Monotonicity}. Clearly the type
of status variation considered in the statement of Axiom \ref{ax:monotonicity}
will change the mean of $u$ and/or the mean of $v$; the type of
status variation considered in Axiom 2$'$ will leave these means
unaltered. The modified version of monotonicity in Axiom 2$'$ will
again ensure that minimal-mobility property is satisfied. Also Axiom
2$'$ is clearly satisfied by the normalised measure.

One consequence of using the second version of the ``More movement,
more mobility'' principle is that it allows for a further step in
normalisation of the mobility measures. It may be appropriate that
the mobility measure remain unchanged under a scale change $\lambda_{0}>0$
in the $0$-distribution and under a scale change $\lambda_{1}>0$
in the $1$-distribution. This strengthens the scale-invariance property
(section \ref{subsec:ordering-scale}) used for mobility orderings
to \emph{scale independence} of the resulting mobility measure. 

Let us examine this development for each of the two classes aggregate
mobility measures derived in section \ref{subsec:aggregate-mobility}. 

\paragraph*{Class-1 mobility measures. }

Setting $\lambda_{0}=1/\mu_{u}$ and $\lambda_{1}=1/\mu_{v}$ it is
clear that (\ref{eq:basic-form-uv-0}) becomes
\begin{equation}
\psi\left(\frac{1}{n}\sum_{i=1}^{n}\left[\frac{u_{i}}{\mu_{u}}\right]^{\alpha}\left[\frac{v_{i}}{\mu_{v}}\right]^{1-\alpha}-\theta\left(1,1\right),1,1\right)=\overline{\psi}\left(\frac{1}{n}\sum_{i=1}^{n}\left[\left[\frac{u_{i}}{\mu_{u}}\right]^{\alpha}\left[\frac{v_{i}}{\mu_{v}}\right]^{1-\alpha}-1\right]\right),\label{eq:basic-form-uv-psi}
\end{equation}
where $\overline{\psi}\left(t\right):=\psi\left(t,1,1\right)$. 

\paragraph*{Class-2 mobility measures. }

Clearly we may obtain a mean-normalised version of (\ref{eq:Mobility-linear})
by dividing $u_{i}$ by $\mu_{u}$ and each $v_{i}$ by $\mu_{v}$
so that
\begin{equation}
d_{i}=\frac{v_{i}}{\mu_{v}}-\frac{u_{i}}{\mu_{u}}.\label{eq:weighted-difference-1}
\end{equation}
and mobility is measured by (\ref{eq:Mobility-linear}) with $d_{i}$
given by (\ref{eq:weighted-difference-1}). Specific examples of these
mean-normalised measures are given in section \ref{sec:discussion}.

\section{Mobility measures: development\label{sec:discussion}}

Expressions (\ref{eq:evaluation_function_PSI+OSI}) and (\ref{eq:DTI=000026PTI-0})
characterise the bases for two classes of mobility measures. Here
we consider the derivation of practical measures from these two bases.

\subsection{Class-1 mobility measures\label{subsec:Class-1-mobility}}

To ensure that the mobility measure $\mathcal{S}_{\alpha}^{1}$ is
well-defined and non-negative for all values of $\alpha$ and that,
for any profile $\mathbf{z}$, $\mathcal{S}_{\alpha}^{1}$ is continuous
in $\alpha$, we adopt the cardinalisation of (\ref{eq:basic-form-uv-psi})
in this definition of \uline{S}cale-independent class-1 mobility
measures: 
\begin{equation}
\mathcal{S}_{\alpha}^{1}:=\begin{cases}
\frac{1}{\alpha\left[\alpha-1\right]n}\sum_{i=1}^{n}\left[\left[\frac{u_{i}}{\mu_{u}}\right]^{\alpha}\left[\frac{v_{i}}{\mu_{v}}\right]^{1-\alpha}-1\right],\  & \alpha\neq0,1,\\
-\frac{1}{n}\sum_{i=1}^{n}\frac{v_{i}}{\mu_{v}}\log\left(\left.\frac{u_{i}}{\mu_{u}}\right/\frac{v_{i}}{\mu_{v}}\right), & \alpha=0,\\
\frac{1}{n}\sum_{i=1}^{n}\frac{u_{i}}{\mu_{u}}\log\left(\left.\frac{u_{i}}{\mu_{u}}\right/\frac{v_{i}}{\mu_{v}}\right) & \alpha=1
\end{cases}\label{eq:mobility-c1-scale}
\end{equation}
$\text{\ensuremath{\mathcal{S}_{\alpha}^{1}}}$ satisfies the second
version of the movement principle (formalised as monotonicity-2, Axiom
2$'$) and constitutes a \textit{class} of aggregate mobility measures
that are independent of population size and independent of the scale
of status.

\subsubsection{Choice of Class-1 mobility measure\label{subsec:Choice-of-mobility-measure} }

An individual member of the mobility class $\mathcal{S}_{\alpha}^{1}$
(\ref{eq:mobility-c1-scale}) is characterised by the choice of $\alpha$.
The choice of $\alpha$ reflects the type of sensitivity that the
user of the mobility measure wishes to incorporate into the measure:
a high positive $\alpha$ produces a measure that is particularly
sensitive to downward movements and a negative $\alpha$ yields a
measure that is sensitive to upward movements.

Furthermore, let us consider a case where every person's upward mobility
is matched by a symmetric downward mobility of someone with the opposite
history ($\forall i,\exists j$ such that $u_{j}=v_{i},v_{j}=u_{i}$).
In this particular case of (perfect) symmetry between downward and
upward status movements, we have $\mu_{u}=\mu_{v}$. Then, it is clear
from (\ref{eq:mobility-c1-scale}) that a high positive $\alpha$
produces a measure that is particularly sensitive to downward movements
(where $u$ exceeds $v$) and a negative $\alpha$ yields a measure
that is sensitive to upward movements (where $v$ exceeds $u$). In
the symmetric downward/upward mobility case, we find
\begin{equation}
\text{\ensuremath{\mathcal{S}_{\alpha}^{1}}}=\frac{1}{\alpha\left[\alpha-1\right]n}\sum_{i=1}^{n}\left(\frac{v_{i}}{\mu_{v}}\left[\frac{u_{i}}{v_{i}}\right]^{\alpha}-1\right).\label{eq:M_alpha_symm}
\end{equation}

\subsubsection{Extending Class-1 mobility measures\label{subsec:Extending-Class-1-mobility}}

We can extend the class (\ref{eq:mobility-c1-scale}) to generate
a different class of mobility measures just by modifying the status
concept. If we replace $u$ and $v$ in (\ref{eq:mobility-c1-scale})
by $u+c$ and $v+c$ where $c$ is a non-negative location parameter,
then (\ref{eq:mobility-c1-scale}) will be replaced by 
\begin{equation}
\frac{\theta\left(c\right)}{n}\sum_{i=1}^{n}\left[\left[\frac{u_{i}+c}{\mu_{u}+c}\right]^{\tilde{\alpha}(c)}\left[\frac{v_{i}+c}{\mu_{v}+c}\right]^{1-\tilde{\alpha}(c)}-1\right],\:\theta\left(c\right)=\frac{1+c^{2}}{\tilde{\alpha}(c)^{2}-\tilde{\alpha}(c)},\:\tilde{\alpha}(c)\neq0,1,\label{eq:mobility-ext}
\end{equation}
where the term $\tilde{\alpha}(c)$ indicates that the sensitivity
parameter may depend upon the location parameter $c$.\footnote{For $\tilde{\alpha}(c)=0$ and $\tilde{\alpha}(c)=1$ there are obvious
special cases of (\ref{eq:mobility-ext}) corresponding to the cases
$\alpha=0,1$ of (\ref{eq:mobility-c1-scale}).}

Take this extension further by writing $\tilde{\alpha}(c)=\gamma+\alpha c$.
Then, taking logs of the expression inside the summation in (\ref{eq:mobility-ext})
we have
\begin{equation}
\log\left(\frac{1+\frac{v}{c}}{1+\frac{\mu_{v}}{c}}\right)+\left[\gamma+\alpha c\right]\left[\log\left(1+\frac{u}{c}\right)+\log\left(1+\frac{\mu_{v}}{c}\right)-\log\left(1+\frac{v}{c}\right)-\log\left(1+\frac{\mu_{u}}{c}\right)\right].\label{eq:log-of-main-expression}
\end{equation}
Using the standard expansion of $\log\left(1+t\right)$ and letting
$c\rightarrow\infty$ we find that, for finite $\alpha,\gamma,\mu_{u},\mu_{v},u,v$,
(\ref{eq:log-of-main-expression}) becomes $\alpha\left[u-\mu_{u}-v+\mu_{v}\right]$
and (\ref{eq:mobility-ext}) becomes
\begin{equation}
\frac{1}{n\alpha^{2}}\sum_{i=1}^{n}\left[\mathrm{e}^{\alpha\left[u_{i}-\mu_{u}-v_{i}+\mu_{v}\right]}-1\right],\label{eq:mobility-limit}
\end{equation}
for any $\alpha\neq0$. The form (\ref{eq:mobility-limit}) gives
the class of mobility measures that would result from using the solution
(\ref{eq:DTI=000026PTI-1}) in the case where DTI and PTI are jointly
imposed on the evaluation function $\phi_{i}$. 

Let $t_{i}=u_{i}-\mu_{u}-v_{i}+\mu_{v}$ and note that $\frac{1}{n}\sum_{i=1}^{n}q_{i}=0$.
Then, using a standard expansion of $\mathrm{e}^{\alpha t_{i}}$,
it can further be shown that, as $\alpha\rightarrow0$, the expression
$\sum_{i=1}^{n}\left[\mathrm{e}^{\alpha t_{i}}-1\right]/\alpha^{2}$
becomes $\frac{1}{2n}\sum_{i=1}^{n}t_{i}^{2}$ . And so the limiting
form of (\ref{eq:mobility-limit}) for $\alpha=0$ is $\frac{1}{2}\text{\textrm{var}}\left(v_{i}-u_{i}\right)$.
This means that the \uline{T}ranslation-independent class-1 mobility
measures are given by:\footnote{For details of the derivation see \citeN{CoFl:18MM},}
\begin{equation}
\mathcal{T}_{\alpha}^{1}:=\begin{cases}
\frac{1}{n\alpha^{2}}\sum_{i=1}^{n}\left[\mathrm{e}^{\alpha\left[u_{i}-\mu_{u}-v_{i}+\mu_{v}\right]}-1\right], & \alpha\neq0,\\
\frac{1}{2}\text{\textrm{var}}\left(v_{i}-u_{i}\right). & \alpha=0.
\end{cases}\label{eq:mobility-c1-trans}
\end{equation}
This class of mobility measures has the property that mobility is
independent of uniform absolute additions to/subtractions from everyone's
status. 

Finally note that, for any positive, finite value of $c$, expression
(\ref{eq:mobility-ext}) would give a class of mobility measures with
properties that are intermediate between the scale-independent measures
(\ref{eq:mobility-c1-scale}) and the translation-independent measures
(\ref{eq:mobility-c1-trans}).

\subsection{Class-2 mobility measures\label{subsec:Class-2-mobility-indices}}

This class of measures focuses on the aggregation of status differences
$d_{i}$ defined in (\ref{eq:simple-difference}). Consider their
behaviour in respect of the two interpretations of the movement principle
(Axiom \ref{ax:monotonicity} and Axiom 2$'$). 

\subsubsection{Non-normalised status\label{subsec:Non-normalised-status}}

Using (\ref{eq:Mobility-linear}) define $i^{*}$ as the largest $i$
such that $d_{(i)}<0$. In order to conform to Axiom \ref{ax:monotonicity}
$a_{i}$ must satisfy $a_{i}<0$ for $i\leq i^{*}$ and $a_{i}\geq0$
otherwise. Consider the simple specification 
\begin{equation}
a_{i}=\begin{cases}
-1 & \text{if }i\leq i^{*}\\
+1 & \text{if }i>i^{*}
\end{cases}.\label{eq:a_i-binary}
\end{equation}
Then mobility is simply (\ref{eq:Mobility-linear}). If status is
income, this becomes the $FO_{1}$ index discussed in section \ref{subsec:ineq-related-measures}
and, if status is log-income, it becomes the $FO_{2}$ index. Furthermore
there is a connection with the \citeANP{BaCa:25} indices in the case
where $\alpha=1$. If status is income and $a_{i}=-1/u_{i}$ when
$i\leq i^{*}$ and 0 otherwise, then (\ref{eq:Mobility-linear}) becomes
$BC_{\mathrm{D}}$; if $a_{i}=1/u_{i}$ when $i>i^{*}$ and 0 otherwise
then (\ref{eq:Mobility-linear}) becomes $BC_{\mathrm{U}}$; if $a_{i}=-1/u_{i}$
when $i\leq i^{*}$ and $1/u_{i}$ otherwise, then (\ref{eq:Mobility-linear})
becomes $BC_{\mathrm{D}}+BC_{\mathrm{U}}$.

However, (\ref{eq:Mobility-linear}) does not fulfil the second interpretation
of  the movement principle. Take the case where for persons $i$ and
$j$, the distances are $d_{i}\geq0$ and $d_{j}\leq0$. Now suppose
that a change occurs to these status levels such that $\Delta d_{i}>0$
and $\Delta d_{j}=-\Delta d_{i}<0$: if a mobility index satisfies
Axiom 2$'$ then mobility must increase with this change -- an increase
in movement; but the index (\ref{eq:Mobility-linear}) remains unchanged.
Clearly this problem could be avoided if (\ref{eq:a_i-binary}) were
replaced by $\phi\left(i-i^{*}-\epsilon\right)$\footnote{Monotonicity requires $a_{i}<0$ if $i\leq i^{*}$. From (\ref{eq:positional-weights-1})
it is clear that if $i^{*}=0$ we have $a_{i}>0$ for $i=1,...,n$;
if $i^{*}=n$ we have $a_{i}<0$ for $i=1,...,n$.} where $\phi$ is an increasing function with $\phi\left(0\right)=0$
and $\epsilon$ is a number between 0 and 1. Setting $\epsilon=\nicefrac{1}{2}$
and normalising $\phi$ so that it is independent of population size
we have

\begin{equation}
a_{i}=\phi\left(\frac{i}{n}-p-\frac{1}{2n}\right),\label{eq:positional-weight}
\end{equation}
where $p$ is the proportion of downward movers, defined as 
\begin{align}
p:=\frac{1}{n}\sum_{i=1}^{n}\mathbbm{1}\left(v_{i}<u_{i}\right).\label{eq:p_fixed}
\end{align}
Clearly (\ref{eq:p_fixed}) implies $p=i^{*}/n$. If $d_{i}$ increases
for any $i>i^{*}$ then measured mobility would increase (in accordance
with the first interpretation of the movement principle); if, for
any $i,j$ where $i>i^{*}>j$ the individual status levels change
such that $\Delta d_{i}>0$ and $\Delta d_{j}=-\Delta d_{i}<0,$ then
again measured mobility would increase (in accordance with the second
interpretation of the movement principle).

An interesting special case of (\ref{eq:positional-weight}) is where
$\phi$ is linear so that 

\begin{equation}
a_{i}=\frac{i}{n}-p-\frac{1}{2n},\label{eq:positional-weights-1}
\end{equation}
which again satisfies Axiom 2$'$ (monotonicity-2) for all mobility
profiles. The associated mobility measure is given by 
\begin{equation}
\frac{1}{n}\sum_{i=1}^{n}\frac{i}{n}d_{(i)}-\left[p+\frac{1}{2n}\right]\mu_{d},\label{eq:Mobility-linear-1}
\end{equation}
where $\mu_{d}$ is the mean of the status differences $d_{i}$. Notice
that (\ref{eq:Mobility-linear-1}) can be rewritten as
\begin{equation}
\nicefrac{1}{2}G+\mu_{d}\left[\frac{1}{2}-p\right],\label{eq:relation-Gamma-G}
\end{equation}
where $G$ is the absolute Gini coefficient of the status differences
$d_{i}$:

\begin{equation}
G:=\frac{2}{n}\sum_{i=1}^{n}\frac{i}{n}d_{(i)}-\mu_{d}\frac{n+1}{n}=\frac{1}{2n^{2}}\sum_{i=1}^{n}\sum_{j=1}^{n}\left|d_{i}-d_{j}\right|,\label{eq:Gini-differences}
\end{equation}
and that, for the case of symmetric mobility, (\ref{eq:relation-Gamma-G})
becomes $\nicefrac{1}{2}G$. This has a particularly nice interpretation
as half the absolute Gini\footnote{Clearly the absolute Gini satisfies Axiom \ref{ax:monotonicity} in
the case of symmetric mobility and satisfies Axiom 2$'$ for all mobility
profiles.} applied to the status differences $d_{i}.$

It is also useful to consider a generalisation of the weights (\ref{eq:positional-weights-1})
as follows

\begin{equation}
a_{i}=\left[\frac{i}{n}-p-\frac{1}{2n}\right]^{\gamma},\label{eq:power_weights}
\end{equation}
which, with (\ref{eq:Mobility-linear}) would give the class-2 \uline{A}bsolute
mobility measures, defined as follows:

\begin{equation}
\mathcal{A}_{\gamma}^{2}:=\frac{1}{n}\sum_{i=1}^{n}\left[\frac{i}{n}-p-\frac{1}{2n}\right]^{\gamma}d_{(i)}\qquad\text{with }\quad d_{i}=v_{i}-u_{i}\label{eq:mobility-c2-abs}
\end{equation}

Clearly an $\mathcal{A}_{\gamma}^{2}$ mobility measure is effectively
a generalised (absolute) Gini applied to the distribution of income
differences $d_{i}.$ The reason why the generalised weights (\ref{eq:power_weights})
are so useful is explained further in section \ref{subsec:Decomposability}
below.

\subsubsection{Mean-normalised status}

Expressions (\ref{eq:Mobility-linear}) and (\ref{eq:Mobility-linear-1})
can be modified to scale-independent versions by replacing the $d_{i}$
in (\ref{eq:simple-difference}) with\footnote{Note that the mean-normalised version of (\ref{eq:Mobility-linear-1})
is not proportional to the conventional Gini evaluated over the weighted
status differences.}

\begin{equation}
d_{i}=\frac{v_{i}}{\mu_{v}}-\frac{u_{i}}{\mu_{u}}.\label{eq:mean-normalised-difference}
\end{equation}

In the case of class-1 mobility measures (in section \ref{subsec:Class-1-mobility}
above), a similar modification immediately gives a class of measures
that satisfy the second interpretation of the movement principle (Axiom
2$'$). However, in the case of these class-2 mobility measures, mean-normalisation
does not change their behaviour in this respect. To see this take,
as before, $i$ and $j$ such that $d_{i}\geq0\geq d_{j}$ and consider
$\Delta v_{i}>0$ and $\Delta v_{j}=-\Delta v_{i}<0$; this will ensure
that $\Delta d_{j}=-\Delta d_{i}<0$ and it is clear that once again
this leads to no change in the mobility measure if the weights $a_{i}$
are given by (\ref{eq:a_i-binary}) and an increase in mobility if
the weights $a_{i}$ are given by (\ref{eq:positional-weights-1}).

The adjusted  concept (\ref{eq:mean-normalised-difference}) and the
generalised weights (\ref{eq:power_weights}) give the \uline{S}cale-independent
class-2 mobility measures:

\begin{equation}
\mathcal{S}_{\gamma}^{2}:=\frac{1}{n}\sum_{i=1}^{n}\left[\frac{i}{n}-p-\frac{1}{2n}\right]^{\gamma}d_{(i)}\qquad\text{with }\quad d_{i}=\frac{v_{i}}{\mu_{v}}-\frac{u_{i}}{\mu_{u}}\label{eq:mobility-c2-scale}
\end{equation}

Scale-independent class-2 mobility measures are translation-independent
only if $\mu_{u}=\mu_{v}.$ To complement $\mathcal{S}_{\gamma}^{2}$,
expressions (\ref{eq:Mobility-linear}) and (\ref{eq:Mobility-linear-1})
can be modified to translation-independent versions by replacing the
$d_{i}$ in (\ref{eq:simple-difference}) with

\begin{equation}
d_{i}=\left[v_{i}-u_{i}\right]-\left[\mu_{v}-\mu_{u}\right].\label{eq:adjusted-distance}
\end{equation}

The adjusted  concept (\ref{eq:adjusted-distance}) and the generalised
weights (\ref{eq:power_weights}) give the \uline{T}ranslation-independent
class-2 mobility measures: 

\begin{equation}
\mathcal{T}_{\gamma}^{2}:=\frac{1}{n}\sum_{i=1}^{n}\left[\frac{i}{n}-p-\frac{1}{2n}\right]^{\gamma}d_{(i)}\qquad\text{with }\quad d_{i}=\left[v_{i}-u_{i}\right]-\left[\mu_{v}-\mu_{u}\right].\label{eq:mobility-c2-trans}
\end{equation}

There is an important issue regarding the weights to be used in definitions
(\ref{eq:mobility-c2-scale}) and (\ref{eq:mobility-c2-trans}): should
the $p$ used in (\ref{eq:power_weights}) be (a) defined by individual
status (whether the condition $v_{i}<u_{i}$ is satisfied), or (b)
defined according to the relevant distance concept used to evaluate
mobility? Option (a) requires that (\ref{eq:p_fixed}) applies in
all cases; option (b) requires that 
\begin{align}
p=\frac{1}{n}\sum_{i=1}^{n}\mathbbm{1}\left(d_{i}<0\right),\label{eq:p_recomp}
\end{align}
where $d_{i}$ is variously given by (\ref{eq:simple-difference}),
(\ref{eq:mean-normalised-difference}) or (\ref{eq:adjusted-distance}).
This issue becomes salient when we consider the decomposition of mobility
in in section \ref{subsec:Decomposability} below.

\subsection{The two classes: overview}

In order to illustrate the properties of mobility measures developed
in sections \ref{subsec:Class-1-mobility} and \ref{subsec:Class-2-mobility-indices}
we compute the values of class-1 and class-2 measures in Table \ref{tab:comparisons_new-1}.
We can see that: 
\begin{itemize}
\item The absolute mobility measures $\mathcal{A}_{\alpha}^{1}$ and $\mathcal{A}_{\alpha}^{2}$
are sensitive to any individual movement in status: they take different
(non-zero) values in all scenarios. 
\item The scale-independent mobility indices $\mathcal{S}_{\alpha}^{1}$
and $\mathcal{S}_{\alpha}^{2}$ are not sensitive to a change in the
status scale in either period: scenarios $1^{c}$ and $1^{d}$ take
the same value and scenario $1^{a}$ takes the value 0. 
\item The translation-independent mobility indices $\mathcal{T}_{\alpha}^{1}$
and $\mathcal{T}_{\alpha}^{2}$ are not sensitive to a change in the
status level in either period: scenarios $1^{c}$ and $1^{e}$ take
the same value and scenario $1^{b}$ takes the value 0. 
\end{itemize}
\begin{table}[tb]
\begin{tabular}{l@{}c@{}cccccccc}
\hline 
 & period &  & \multicolumn{7}{c}{period}\tabularnewline
\cline{4-10} &  &  &  &  &  &  &  &  & \tabularnewline
 & 0 &  & $1^{a}$ & $1^{b}$ & $1^{c}$ & $1^{d}$ & $1^{e}$ & $1^{f}$ & $1^{g}$\tabularnewline
\hline &  &  &  &  &  &  &  &  & \tabularnewline
A & $10$ &  & $20$ & $15$ & $20$ & $40$ & $25$ & $10$ & $10$\tabularnewline
B & $20$ &  & $40$ & $25$ & $40$ & $80$ & $45$ & $30$ & $40$\tabularnewline
C & $40$ &  & $80$ & $45$ & $10$ & $20$ & $15$ & $40$ & $160$\tabularnewline
\hline &  &  &  &  &  &  &  &  & \tabularnewline
$\mathcal{A}_{\alpha}^{1}$ in (\ref{eq:mobility-c1-abs}) &  &  & 32.347 & 5.654 & 9.242 & 50.831 & 14.896 & 4.055 & 83.178\tabularnewline
$\mathcal{A}_{\gamma}^{2}$ in (\ref{eq:mobility-c2-abs}) &  &  & 15 & 2.5 & 5.556 & 12.778 & 6.389 & 2.778 & 36.667\tabularnewline
$\mathcal{S}_{\alpha}^{1}$ in (\ref{eq:mobility-c1-scale}) &  &  & 0 & 0.005 & 0.396 & 0.396 & 0.332 & 0.019 & 0.090\tabularnewline
$\mathcal{S}_{\gamma}^{2}$ in (\ref{eq:mobility-c2-scale}) &  &  & 0 & 0.025 & 0.238 & 0.238 & 0.213 & 0.054 & 0.095\tabularnewline
$\mathcal{T}_{\alpha}^{1}$ in (\ref{eq:mobility-c1-trans}) &  &  & 116.667 & 0 & 350 & 816.667 & 350 & 16.667 & 2066.667\tabularnewline
$\mathcal{T}_{\gamma}^{2}$ in (\ref{eq:mobility-c2-trans}) &  &  & 3.333 & 0 & 5.556 & 8.889 & 5.556 & 1.111 & 13.333\tabularnewline
\hline 
\end{tabular}\caption{Mobility measures in different scenarios, with $\alpha=0$ and $\gamma=1$.}
\label{tab:comparisons_new-1}
\end{table}

\subsection{Decomposability\label{subsec:Decomposability}}

\subsubsection{Class-1 mobility measures}

The axioms underpinning Class-1 mobility measures induce an additive
structure for the mobility index, so that the mobility measures $\mathcal{S}_{\alpha}^{1}$
in (\ref{eq:mobility-c1-scale}) are clearly decomposable by arbitrary
population subgroups.

Let there be $K$ groups and let the proportion of population falling
in group $k$ be $p_{k}$. Then the class of absolute mobility measures
(\ref{eq:mobility-c1-abs}) can be expressed as $\mathcal{A}_{\alpha}^{1}=\sum_{k=1}^{K}p_{k}\mathcal{A}_{\alpha,k}^{1}$.

The class of scale-independent mobility measures (\ref{eq:mobility-c1-scale})
can be expressed as: 
\begin{equation}
\mathcal{S}_{\alpha}^{1}=\sum_{k=1}^{K}p_{k}\left[\frac{\mu_{u,k}}{\mu_{u}}\right]^{\alpha}\left[\frac{\mu_{v,k}}{\mu_{v}}\right]^{1-\alpha}\mathcal{S}_{\alpha,k}^{1}+\frac{1}{\alpha^{2}-\alpha}\left[\sum_{k=1}^{K}p_{k}\left[\frac{\mu_{u,k}}{\mu_{u}}\right]^{\alpha}\left[\frac{\mu_{v,k}}{\mu_{v}}\right]^{1-\alpha}-1\right]\label{eq:decomp}
\end{equation}
for $\alpha\not=0,1$, where $\mu_{u,k}$ ($\mu_{v,k}$) is the mean
status in period-0 (period-1) in group $k$, and $\mu_{u},\mu_{v}$
are the corresponding population means (so that $\mu_{u}=K^{-1}\sum_{k=1}^{K}p_{k}\mu_{u,k}$,
$\mu_{v}=K^{-1}\sum_{k=1}^{K}p_{k}\mu_{v,k}$). In particular, notice
that in the case where $u=x$ and $v=\mu{}_{x}$, we obtain the standard
formula of decomposability for the class of GE inequality measures
\cite{Cowe:11MI}. We have the following limiting forms for the cases
$\alpha=0$ and $\alpha=1$, respectively 
\begin{equation}
\mathcal{S}_{0}^{1}=\sum_{k=1}^{K}p_{k}\left[\frac{\mu_{v,k}}{\mu_{v}}\right]\mathcal{S}_{0}^{1}{}_{,k}-\sum_{k=1}^{K}p_{k}\left[\frac{\mu_{v,k}}{\mu_{v}}\right]\log\left(\left.\frac{\mu_{u,k}}{\mu_{u}}\right/\frac{\mu_{v,k}}{\mu_{v}}\right)
\end{equation}
\begin{equation}
\mathcal{S}_{1}^{1}=\sum_{k=1}^{K}p_{k}\left[\frac{\mu_{u,k}}{\mu_{u}}\right]\mathcal{S}_{1}^{1}{}_{,k}+\sum_{k=1}^{K}p_{k}\left[\frac{\mu_{u,k}}{\mu_{u}}\right]\log\left(\left.\frac{\mu_{u,k}}{\mu_{u}}\right/\frac{\mu_{v,k}}{\mu_{v}}\right)
\end{equation}

This means, for example, that we may partition the population unambiguously
into an upward status group $\mathsf{U}$ (for $u_{i}\leq v_{i}$)
and a downward status group $\mathsf{D}$ (for $u_{i}>v_{i}$) and,
using an obvious notation, express overall mobility as 
\begin{equation}
\mathcal{S}_{\alpha}^{1}=w^{\mathsf{U}}\mathcal{S}_{\alpha}^{1\mathsf{U}}+w^{\mathsf{D}}\mathcal{S}_{\alpha}^{1\mathsf{D}}+\mathcal{S}_{\alpha}^{1\mathsf{btw}},\label{eq:decompUD}
\end{equation}
where the weights $w^{\mathsf{U}}$, $w^{\mathsf{D}}$ and the between-group
mobility component $\mathcal{S}_{\alpha}^{1\mathsf{btw}}$ are functions
of the status-means for each of the two groups and overall; comparing
$\mathcal{S}_{\alpha}^{1\mathsf{U}}$ and $\mathcal{S}_{\alpha}^{1\mathsf{D}}$
enables one to say precisely where mobility has taken place -- see
also \citeN{RaGe:23}, equation (27) and \citeN{BaCa:18}.

We may use this analysis to extend the discussion of the choice of
$\alpha$ in section \ref{subsec:Class-1-mobility}. Suppose we create
a ``reverse profile'' $\mathbf{z'\left(z\right)}:=\left\{ z'_{i}=\left(v_{i},u_{i}\right)|\;z_{i}=\left(u_{i},v_{i}\right),i=1,...,n\right\} $
by reversing each person's history -- swapping the $u$s and $v$s
in (\ref{eq:mobility-c1-scale}); then we have $\mathcal{S}_{\alpha}^{1}\mathbf{(z'\left(z\right)})=\mathcal{S}_{1-\alpha}^{1}(\mathbf{z})$.
So, consider a special case of the upward and downward status groups
$\mathsf{U}$ and $\mathsf{D}$ where each member of $\mathsf{U}$
is matched by a member of $\mathsf{D}$ with the reverse history --
and vice versa. In this case we would have 
\begin{equation}
\mathcal{S}_{\alpha}^{1\mathsf{U}}=\mathcal{S}_{1-\alpha}^{1\mathsf{D}}.\label{eq:decompUDa}
\end{equation}
It suggests that mobility measurement of upward movements and of symmetric
downward movements would be identical with $\alpha=0.5$ ($\mathcal{S}_{0.5}^{\mathsf{1U}}=\mathcal{S}_{0.5}^{\mathsf{1D}}$).
Furthermore, mobility measurement of upward movements with $\alpha=1$
would be identical to mobility measurement of symmetric downward movements
with $\alpha=0$ ($\mathcal{S}_{1}^{\mathsf{1U}}=\mathcal{S}_{0}^{\mathsf{1D}}$).

In the mobility index $\mathcal{S}_{\alpha}^{1}$, the weights given
to upward mobility and to downward mobility can be studied through
its decomposability property. With symmetric upward/downward status
movements, from (\ref{eq:decomp}) and (\ref{eq:decompUD}), we can
see that\footnote{From (\ref{eq:decomp}) and (\ref{eq:decompUD}), we have $w^{\mathsf{U}}=p_{1}(\mu_{u,1}/\mu_{u})^{\alpha}(\mu_{v,1}/\mu_{v})^{1-\alpha}$
and $w^{\mathsf{D}}=p_{2}(\mu_{u,2}/\mu_{u})^{\alpha}(\mu_{v,2}/\mu_{v})^{1-\alpha}$.
With symmetric downward/upward mobility, we also have $p_{1}=p_{2}$,
$\mu_{u,1}=\mu_{v,2}<\mu_{v,1}=\mu_{u,2}$ and $\mu_{u}=\mu_{v}$.
Then, $w^{\mathsf{U}}/w^{\mathsf{D}}=(\mu_{u,2}/\mu_{v,2})^{1-2\alpha}$,
which is greater (less) than one if $1-2\alpha>(<)0$. } 
\begin{enumerate}
\item for $\alpha=0.5$, we have $w^{\mathsf{U}}=w^{\mathsf{D}}$, 
\item for $\alpha<0.5$, we have $w^{\mathsf{U}}>w^{\mathsf{D}}$, 
\item for $\alpha>0.5$, we have $w^{\mathsf{U}}<w^{\mathsf{D}}$. 
\end{enumerate}
In other words, $\alpha=0.5$ puts the same weight on both upward
and downward mobility components in (\ref{eq:decomp}), while $\alpha<0.5$
($\alpha>0.5$) puts more weights on upward (downward) mobility component.
The sensitivity parameter $\alpha$ enables us to capture \emph{directional
sensitivity} in the mobility context:\footnote{See also : \citeN{BhMa:11}, \shortciteN{CoLiMa:14}, \shortciteN{DeVd:10}
and \shortciteN{ScVdg:11}.} high positive values result in a mobility index that is more sensitive
to downward movements from period 0 to period 1; negative $\alpha$
is more sensitive to upward movements. Picking a value for this parameter
is a normative choice.

We also have:%

\textbf{
\begin{equation}
\mathcal{T}_{\alpha}^{1}=\sum_{k=1}^{K}w_{k}\mathcal{T}_{\alpha,k}^{1}+\mathcal{T}_{\alpha}^{1\mathsf{\mathsf{Btw}}}
\end{equation}
}

where $w_{k}:=\frac{n_{k}}{n}\mathrm{e}^{\alpha\left[\mu_{u,k}-\mu_{u}-\mu_{v,k}+\mu_{v}\right]}$
and 

\begin{equation}
\mathcal{T}_{\alpha}^{1\mathsf{\mathsf{Btw}}}=\begin{cases}
\frac{1}{\alpha^{2}}\left[\sum_{k=1}^{K}w_{k}-1\right] & \alpha\neq0,\\
\frac{1}{2}\left[\sum_{k=1}^{K}w_{k}\mu_{d,k}^{2}-\mu_{d}^{2}\right] & \alpha=0
\end{cases}.
\end{equation}

\subsubsection{Class-2 mobility measures: Upward and Downward }

As with inequality measures such as the Gini coefficient, exact decomposition
of mobility by population subgroups is not usually possible. But for
Upward/Downward decompositions of the non-normalised mobility measures
in section \ref{subsec:Class-2-mobility-indices} we have easily interpretable
results. Clearly (\ref{eq:Mobility-linear}) can be rewritten as
\begin{eqnarray}
-p\frac{1}{np}\sum_{i=1}^{np}d_{(i)}+\left[1-p\right]\frac{1}{n-np}\sum_{i=1}^{n-np}d_{(i+np)} & = & -pd^{\mathsf{D}}+\left[1-p\right]d^{\mathsf{U}},\label{eq:decomp_d}
\end{eqnarray}
where $d^{\mathsf{D}}$, $d^{\mathsf{U}}$ are the average weighted
distance of downward and upward moves, respectively. 
\begin{figure}
\noindent \centering{}\includegraphics[scale=0.8]{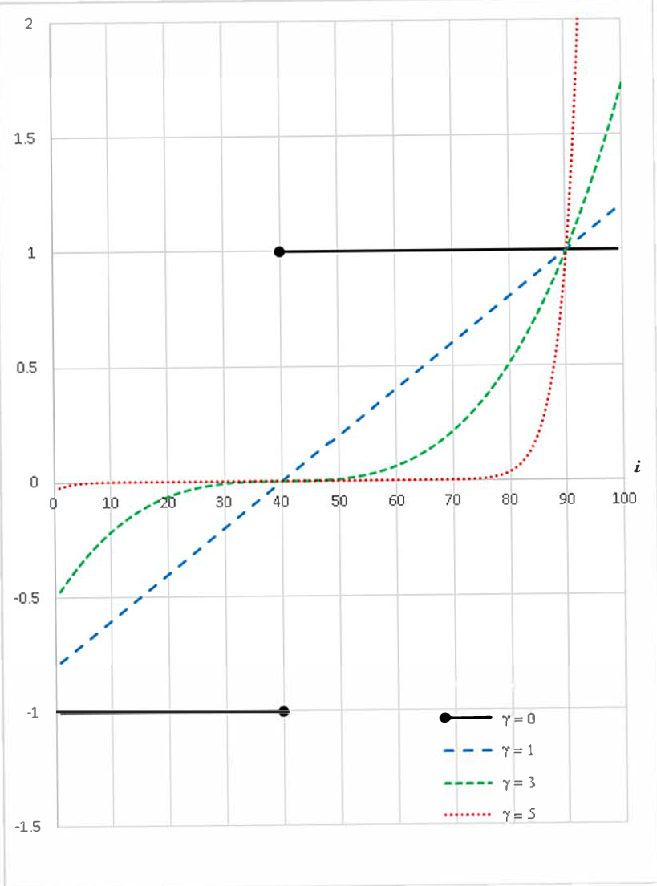}\caption{\label{fig:weights-gamma}The weights $a_{i}$ for different values
of $\gamma$}
\end{figure}

Consider the general class of measures that come from using the weights
(\ref{eq:positional-weight}) in (\ref{eq:simple-difference}): 

\begin{equation}
\Gamma:=\frac{1}{n}\sum_{i=1}^{n}\phi\left(\frac{i}{n}-p-\frac{1}{2n}\right)d_{(i)}\label{eq:Gamma-phi}
\end{equation}
If the function $\phi$ has the property that $\phi\left(\lambda\theta\right)=\psi\left(\lambda\right)\phi\left(\theta\right)$
for all real numbers $\lambda,\theta$ and some function $\psi$ then
(\ref{eq:Gamma-phi}) can be written as\footnote{To see this note that 
\begin{eqnarray*}
\Gamma=p\psi\left(p\right)\frac{1}{np}\sum_{i=1}^{np}\phi\left(\frac{1}{p}\left[\frac{i}{n}-p-\frac{1}{2n}\right]\right)d_{(i)}+\left[1-p\right]\psi\left(1-p\right)\frac{1}{n-np}\sum_{i=np+1}^{n}\phi\left(\frac{1}{1-p}\left[\frac{i}{n}-p-\frac{1}{2n}\right]\right)d_{(i)},\\
\Gamma^{\mathsf{D}}=\frac{1}{np}\sum_{i=1}^{np}\phi\left(\frac{1}{p}\left[\frac{i}{n}-p-\frac{1}{2n}\right]\right)d_{(i)},\\
\Gamma^{\mathsf{U}}=\frac{1}{n-np}\sum_{i=np+1}^{n}\phi\left(\frac{1}{1-p}\left[\frac{i}{n}-p-\frac{1}{2n}\right]\right)d_{(i)}.
\end{eqnarray*}
 } 
\begin{equation}
\Gamma=p\psi\left(p\right)\Gamma^{\mathsf{D}}+\left[1-p\right]\psi\left(1-p\right)\Gamma^{\mathsf{U}}\label{eq:Gamma-decomposition}
\end{equation}
where $\Gamma^{\mathsf{D}}$ and $\Gamma^{\mathsf{U}}$are, respectively,
mobility measured over the set of downward movers, and mobility measured
over the rest of the population using (\ref{eq:simple-difference}): 

\begin{equation}
\Gamma^{\mathsf{D}}=\frac{1}{np}\sum_{i=1}^{np}\phi\left(\frac{i}{np}-1-\frac{1}{2np}\right)d_{(i)},\;\Gamma^{\mathsf{U}}=\frac{1}{n-np}\sum_{i=1}^{n-np}\phi\left(\frac{i}{n-np}-\frac{1}{2\left[n-np\right]}\right)d_{(i+np)}.
\end{equation}

The property (\ref{eq:Gamma-decomposition}) requires that $\phi$
take the form $\phi\left(\theta\right)=A\theta^{\gamma}$ where $A$
and $\gamma$ are constants. This means that the general positional
weights (\ref{eq:positional-weight}) become the power weights (\ref{eq:power_weights});
and the mobility measure becomes $\mathcal{A}_{\gamma}^{2}$ defined
in (\ref{eq:mobility-c2-abs}), where $\gamma$ is a positive integer.\footnote{If $\gamma$ were negative then the weights in (\ref{eq:power_weights})
would go to $\pm\infty$ when $i$ approaches $np$. If $\gamma$
were not an integer then the weights for $i<np+\frac{1}{2}$ would
not exist.} The positive integer $\gamma$ needs to be an odd number. When $\gamma$
is odd the weights in (\ref{eq:power_weights}) increase with $i$
everywhere; they are negative if $i<np+\frac{1}{2}$, positive if
$i>np+\frac{1}{2}$ and 0 if $i=np+\frac{1}{2}$ and so conform to
Axiom \ref{ax:monotonicity}.\footnote{See section \ref{subsec:Non-normalised-status}. If $\gamma$ were
even then the weights would be non-negative everywhere and so would
violate Axiom \ref{ax:monotonicity} for $i<np+\frac{1}{2}$.} Figure \ref{fig:weights-gamma} shows the system of weights that
emerge in a population of 100 where 40 persons experience downward
movement. It is clear that the higher value of $\gamma$ puts more
weight toward the extremes of the distribution of distances $d$.

Consider now the evaluation of mobility using (\ref{eq:Gamma-decomposition})
with power weights (\ref{eq:power_weights}) and a distance function
that could be (\ref{eq:simple-difference}), (\ref{eq:mean-normalised-difference})
or (\ref{eq:adjusted-distance}). Given the use of power weights (\ref{eq:power_weights}),
the evaluation of $\Gamma^{\mathsf{D}}$ and $\Gamma^{\mathsf{U}}$
becomes%
{} 
\begin{equation}
\Gamma^{\mathsf{D}}=\frac{1}{np^{\gamma+1}}\sum_{i=1}^{np}a_{i}d_{(i)}^{\mathsf{D}},\;\Gamma^{\mathsf{U}}=\frac{1}{n\left[1-p\right]{}^{\gamma+1}}\sum_{i=1}^{n-np}a_{i}d_{(i+np)}^{\mathsf{U}}.\label{eq:Gamma_D_and_U}
\end{equation}
where $d_{(i)}^{\mathsf{D}}$ ($d_{(i)}^{\mathsf{U}}$) denotes the
distance evaluated for a person $i$ among the downward-movers (rest
of the population).%
{} Therefore 
\[
p^{\gamma+1}\Gamma^{\mathsf{D}}+\left[1-p\right]{}^{\gamma+1}\Gamma^{\mathsf{U}}=\frac{1}{n}\sum_{i=1}^{np}a_{i}d_{(i)}^{\mathsf{D}}+\frac{1}{n}\sum_{i=np+1}^{n}a_{i}d_{(i-np)}^{\mathsf{U}}.
\]
Rearranging the right-hand side of this we have 

\begin{align}
p^{\gamma+1}\Gamma^{\mathsf{D}}+\left[1-p\right]{}^{\gamma+1}\Gamma^{\mathsf{U}} & =\Gamma+\frac{1}{n}\sum_{i=n}^{n}a_{i}\left[d_{(i)}^{\mathsf{DU}}-d_{(i)}\right],\label{eq:p^(gamma+1)Gamma^D}
\end{align}

\begin{align}
d^{\mathsf{DU}} & :=\left\{ d_{(1)}^{\mathsf{D}},\dots,d_{(np)}^{\mathsf{D}},d_{(1)}^{\mathsf{U}},\dots,d_{(n-np)}^{\mathsf{U}}\right\} ,\label{eq:d^DU}
\end{align}
so that one has the decomposition formula:

\begin{equation}
\Gamma=p^{\gamma+1}\Gamma^{\mathsf{D}}+\left[1-p\right]{}^{\gamma+1}\Gamma^{\mathsf{U}}+\Gamma^{\mathsf{Btw}},\label{eq:decomposition_gamma2}
\end{equation}
with $\Gamma^{\mathsf{Btw}}$, the between-group component, given
by 
\begin{equation}
\Gamma^{\mathsf{Btw}}=\frac{1}{n}\sum_{i=n}^{n}a_{i}\left[d_{(i)}-d_{(i)}^{\mathsf{DU}}\right].\label{eq:decomposition_gamma_btw2}
\end{equation}
It is clear from (\ref{eq:decomposition_gamma_btw2}) that the between-group
component is $\left(d^{\mathsf{DU}},d\right)$, the mobility implied
by moving from the distribution of $d^{\mathsf{DU}}$ to the distribution
of $d$; clearly this quantity will depend on the definition of the
distance $d_{i}$.

The expressions (\ref{eq:decomposition_gamma2}, \ref{eq:decomposition_gamma_btw2})
can then be used immediately to yield decompositions for each of the
mobility indices $\mathcal{A}_{\gamma}^{2}$, $\mathcal{S}_{\gamma}^{2}$
and $\mathcal{T}_{\gamma}^{2}$ on the assumption that $p$ used in
the power weights (\ref{eq:power_weights}) is status-based (see equation
\ref{eq:p_fixed}) and therefore is the same for each of the distance
concepts used in each of these three mobility measures. The decompositions
are as follows: 
\begin{eqnarray}
\mathcal{A}_{\gamma}^{2} & = & p^{\gamma+1}\mathcal{A}_{\gamma}^{2\mathsf{D}}+\left[1-p\right]{}^{\gamma+1}\mathcal{A}_{\gamma}^{2\mathsf{\mathsf{U}}}\label{eq:A2-decomp-gamma}\\
\mathcal{S}_{\gamma}^{2} & = & p^{\gamma+1}\mathcal{S}_{\gamma}^{2\mathsf{D}}+\left[1-p\right]{}^{\gamma+1}\mathcal{S}_{\gamma}^{2\mathsf{U}}+\mathcal{S}^{\mathsf{Btw}}\\
\mathcal{T}_{\gamma}^{2} & = & p^{\gamma+1}\mathcal{T}_{\gamma}^{2\mathsf{D}}+\left[1-p\right]{}^{\gamma+1}\mathcal{T}_{\gamma}^{2\mathsf{U}}+\mathcal{T}^{\mathsf{Btw}}\label{eq:T2-decomp-gamma}
\end{eqnarray}
In the case of $\mathcal{A}_{\gamma}^{2}$, absolute mobility, the
between-group component is zero;\footnote{The reason is that, in this case, $d_{(i)}^{\mathsf{D}}=d_{(i)},i=1,...,np$
and $d_{(i)}^{\mathsf{U}}=d_{np+i},i=1,...,n-np$ so that $d^{\mathsf{DU}}=d$
.} $\mathcal{S}^{\mathsf{Btw}}$ is a scale-independent mobility measure
of $\left(d^{\mathsf{DU}},d\right)$ using (\ref{eq:mean-normalised-difference});
$\mathcal{T}^{\mathsf{Btw}}$ is a translation-independent measure
of $\left(d^{\mathsf{DU}},d\right)$ using (\ref{eq:adjusted-distance}).

Now consider the case where $p$ is defined by the distance concept,
as in equation (\ref{eq:p_recomp}) rather than by status as in equation
(\ref{eq:p_fixed}). In this case we have 
\begin{align}
\Gamma^{\mathsf{D}}=\frac{1}{np}\sum_{i=1}^{np}a_{i}^{\mathsf{D}}d_{(i)}^{\mathsf{D}}\qquad\text{and}\qquad\Gamma^{\mathsf{U}}=\frac{1}{n\left[1-p\right]}\sum_{i=1}^{n-np}a_{i}^{\mathsf{U}}d_{(i)}^{\mathsf{U}},\label{eq:Gamma_D,Gamma_U}
\end{align}
and so
\begin{align}
p\Gamma^{\mathsf{D}}+\left[1-p\right]\Gamma^{\mathsf{U}} & =\frac{1}{n}\sum_{i=1}^{np}a_{i}^{\mathsf{D}}d_{(i)}^{\mathsf{D}}+\frac{1}{n}\sum_{i=1}^{n-np}a_{i}^{\mathsf{U}}d_{(i)}^{\mathsf{U}}.\label{eq:decomposition1b}
\end{align}
In (\ref{eq:Gamma_D,Gamma_U}) and (\ref{eq:decomposition1b}) $a_{i}^{\mathsf{D}},a_{i}^{\mathsf{U}}$
are the weights evaluated for the two groups $\mathsf{D}$ and $\mathsf{U}$
separately; these two groups are is determined by $d_{(i)}$ using
$\mu_{d}$, the mean distance for the whole population. There may
be some $i$ in group $\mathsf{D}$ for whom the movement would appear
to be upwards if the criterion used were the group-specific $d_{(i)}^{\mathsf{D}}$
(which uses the group mean $\mu_{d}^{\mathsf{D}})$; a similar ambiguity
exists n $\mathsf{U}$. So, define the proportion of downward movements
in each of the two groups $\mathsf{D}$ and $\mathsf{U}$ using the
respective group means as follows: 
\begin{align}
p^{\mathsf{D}}=\frac{1}{np}\sum_{i=1}^{np}\mathbbm{1}\left(d_{(i)}^{\mathsf{D}}<0\right)\qquad\text{and}\qquad p^{\mathsf{U}}=\frac{1}{n-np}\sum_{i=1}^{n-np}\mathbbm{1}\left(a_{i}^{\mathsf{U}}<0\right)
\end{align}
Deriving the coefficients $a_{i}^{\mathsf{D}}$ and $a_{i}^{\mathsf{U}}$
presents difficulties for arbitrary $\gamma$, but the special case
$\gamma=1$ is tractable. In this case one has: 
\begin{align}
a_{i}^{\mathsf{D}} & =\frac{1}{p}a_{i}+1-p^{\mathsf{D}}\qquad & i & =1,\dots,np\\
a_{i}^{\mathsf{U}} & =\frac{1}{1-p}a_{i}-p^{\mathsf{U}}\qquad & i & =np+1,\dots,n
\end{align}
Using these in (\ref{eq:decomposition1b}) yields:%
{} 
\begin{align}
p^{2}\Gamma^{\mathsf{D}}+\left[1-p\right]^{2}\Gamma^{\mathsf{U}} & =\Gamma+\frac{1}{n}\sum_{i=n}^{n}a_{i}\left[d_{(i)}^{\mathsf{DU}}-d_{(i)}\right]+p^{2}\left[1-p^{\mathsf{D}}\right]\mu_{d}^{\mathsf{D}}-\left[1-p\right]{}^{2}p^{\mathsf{U}}\mu_{d}^{\mathsf{U}}\label{eq:p^2GammaD}
\end{align}
with $d^{\mathsf{DU}}$ defined in (\ref{eq:d^DU}). Equation (\ref{eq:p^2GammaD})
can be further simplified because, for mean-normalised distance $d_{i}$
-- either the scale-independent version (\ref{eq:mean-normalised-difference})
or the translation-independent version (\ref{eq:adjusted-distance})
-- one finds $\mu_{d}=\mu_{d}^{\mathsf{D}}=\mu_{d}^{\mathsf{U}}=0$,\footnote{To see this, take the mean of $d_{i}$ in equations (\ref{eq:mean-normalised-difference})
and (\ref{eq:adjusted-distance}).} and for the case where $d_{i}$ is given by (\ref{eq:simple-difference})
one has $p^{\mathsf{D}}=1$ and $p^{\mathsf{U}}=0$. Therefore (\ref{eq:p^2GammaD})
reduces to (\ref{eq:p^(gamma+1)Gamma^D}) with $\gamma=1$, and so
we again get the decomposition of $\Gamma$ given by (\ref{eq:decomposition_gamma2},\ref{eq:decomposition_gamma_btw2}).

The net result is that $\mathcal{A}_{1}^{2}$, $\mathcal{S}_{1}^{2}$
and $\mathcal{T}_{1}^{2}$ (the class-2 mobility indices for $\gamma=1$)
with $p$ defined by the distance concept have exactly the same form
as their counterparts with $p$ defined by status -- equations (\ref{eq:A2-decomp-gamma}-\ref{eq:T2-decomp-gamma})
with parameter $\gamma=1$.

\subsubsection{Class-2 mobility measures: structural mobility, exchange mobility
and growth}

Finally we consider an alternative decomposition method that is particularly
convenient when applied to Class-2 mobility measures. This is a way
of breaking down overall mobility into the three components mentioned
in the section title. To implement this form of decomposition we change
the picture of the mobility process from the single-step $u\rightarrow v$
to the two-step $u\rightarrow u'\rightarrow v$, where $u'$ is an
artificially created vector of status levels defined as follows: $u':=\left\{ (u_{1},u_{2},\dots,u_{n})\left|\:\text{rank}(u')=\text{rank}(v)\right.\right\} $
is the set of $u$-values reordered such that individuals have the
same ranks in $v$ and $u'$.

Let us use the symbols $d^{\mathsf{A}},d^{\mathsf{S}},d_{i}^{\mathsf{T}}$
to denote the three separate distance concepts in (\ref{eq:simple-difference}),
(\ref{eq:mean-normalised-difference}), (\ref{eq:adjusted-distance}),
respectively. It is important to note that the proportion of downward
movements $p$ is defined relative to the relevant distance concept.
So, restricting attention to the case $\gamma=1$, the absolute, scale-independent
and translation-independent measures are:

\begin{eqnarray}
\mathcal{A}_{1}^{2}\left(u,v\right) & = & \frac{1}{n}\sum_{i=1}^{n}a_{i}^{\mathsf{A}}d_{(i)}^{\mathsf{A}},\qquad\text{with }\quad a_{i}^{\mathsf{A}}=\frac{i}{n}-p\left(d^{\mathsf{A}}\right)-\frac{1}{2n}\label{eq:mobility-c2-abs-1}\\
\mathcal{S}_{1}^{2}\left(u,v\right) & = & \frac{1}{n}\sum_{i=1}^{n}a_{i}^{\mathsf{S}}d_{(i)}^{\mathsf{S}}\qquad\text{with }\quad a_{i}^{\mathsf{S}}=\frac{i}{n}-p\left(d^{\mathsf{S}}\right)-\frac{1}{2n}\\
\mathcal{T}_{1}^{2}\left(u,v\right) & = & \frac{1}{n}\sum_{i=1}^{n}a_{i}^{\mathsf{T}}d_{(i)}^{\mathsf{T}}\qquad\text{with }\quad a_{i}^{\mathsf{T}}=\frac{i}{n}-p\left(d^{\mathsf{T}}\right)-\frac{1}{2n}\label{eq:mobility-c2-trans-1}
\end{eqnarray}
 We also need to introduce the counterparts $\mathcal{A}_{1}^{2}\left(u',v\right),\mathcal{S}_{1}^{2}\left(u',v\right),\mathcal{T}_{1}^{2}\left(u',v\right)$
for the measurement of the mobility from $u'$ to $v$, where there
is no reranking of individuals; the relevant distance concepts in
these cases are denoted $d^{\mathsf{A'}},d^{\mathsf{S'}},d^{\mathsf{T'}}$
, and the weights $a_{i}^{\mathsf{A'}},a_{i}^{\mathsf{S'}},a_{i}^{\mathsf{T'}}$
are defined according to these modified distance concepts. 

We then find the following simple relationship for each of the Class-2
mobility measures in (\ref{eq:mobility-c2-abs-1})-(\ref{eq:mobility-c2-trans-1}):\footnote{Proofs are in the Appendix.}
\begin{align}
\mathcal{A}_{1}^{2}\left(u,v\right) & =\underset{\mathrm{structural}}{\underbrace{\mathcal{A}_{1}^{2}\left(u',v\right)}}+\underset{\mathrm{exchange}}{\underbrace{\frac{1}{n}\sum_{i=1}^{n}a_{i}^{\mathsf{A'}}\left[d_{(i)}^{\mathsf{A}}-d_{(i)}^{\mathsf{A'}}\right]}}+\underset{\mathrm{growth}}{\underbrace{\left[p\left(d^{\mathsf{A'}}\right)-p\left(d^{\mathsf{A}}\right)\right]\left[\mu_{v}-\mu_{u}\right]}},\\
\mathcal{S}_{1}^{2}\left(u,v\right) & =\underset{\mathrm{structural}}{\underbrace{\mathcal{S}_{1}^{2}\left(u',v\right)}}+\underset{\mathrm{exchange}}{\underbrace{\frac{1}{n}\sum_{i=1}^{n}a_{i}^{\mathsf{S'}}\left[d_{(i)}^{\mathsf{S}}-d_{(i)}^{\mathsf{S'}}\right]}},\label{eq:Structural-Exchange_S}\\
\mathcal{T}_{1}^{2}\left(u,v\right) & =\underset{\mathrm{structural}}{\underbrace{\mathcal{T}_{1}^{2}\left(u',v\right)}}+\underset{\mathrm{exchange}}{\underbrace{\frac{1}{n}\sum_{i=1}^{n}a_{i}^{\mathsf{T'}}\left[d_{(i)}^{\mathsf{T}}-d_{(i)}^{\mathsf{T'}}\right]}}.\label{eq:Structural-Exchange_T}
\end{align}

So, the simple structure of class-2 mobility indices allows a functional
decomposition of overall mobility into two or three components. In
the case of scale-independent ($\mathcal{S}_{1}^{2}$) or translation-independent
($\mathcal{T}_{1}^{2}$) measures -- equations (\ref{eq:Structural-Exchange_S})
and (\ref{eq:Structural-Exchange_T}) -- there are just two components,
\emph{structural} and \emph{exchange}, where the structural component
refers to the mobility involved it in changes in the marginal distributions,
but without changes in individual position.\footnote{The structural-exchange distinction follows from the usage by\textsl{\emph{
}}{\small{}\citeN{VanK:04}}: ``the \emph{structural} contribution
is the share of mobility that can be explained by the evolution of
the shape of the distribution, and the \emph{exchange} contribution
is the proportion of mobility that is due to the reranking of individuals
over the positions available in the economy{\small{}.''}} In the case of absolute mobility $\mathcal{A}_{1}^{2}$ these two
components are supplemented by a third: the \emph{growth} component
comes from the difference in means between $u$ and $v$, weighted
by the changes in the proportion of downward movers as a result of
reranking.

\subsection{Mobility and inequality\label{subsec:Mobility-and-inequality}}

Take the case where the destination is an equal distribution with
the same mean as the origin distribution: 
\begin{equation}
\forall i:v_{i}=\mu_{u}.\label{eq:equal_destination}
\end{equation}
Consider what happens to the two main classes of mobility measures
developed in sections \ref{subsec:Class-1-mobility} and \ref{subsec:Class-2-mobility-indices}.

\paragraph*{Class-1 mobility measures}

We use the class-1 scale-independent mobility indices $\text{\ensuremath{\mathcal{S}_{\alpha}^{1}}}$
defined in (\ref{eq:mobility-c1-scale}). If we posit the formulation
in (\ref{eq:equal_destination}) for the destination distribution
then (\ref{eq:mobility-c1-scale}) become the generalised entropy
class of inequality measures {\small{}\cite{CoFl:21IM}} 
\begin{equation}
\left\{ \begin{array}{ccc}
\frac{1}{\alpha\left[\alpha-1\right]n}\sum_{i=1}^{n}\left[\left[\frac{u_{i}}{\mu_{u}}\right]^{\alpha}-1\right], &  & \text{\ensuremath{\alpha\neq}0,1,}\\
\\
-\frac{1}{n}\sum_{i=1}^{n}\log\left(\frac{u_{i}}{\mu_{u}}\right), &  & \text{\ensuremath{\alpha=}0,}\\
\\
\frac{1}{n}\sum_{i=1}^{n}\frac{u_{i}}{\mu_{u}}\log\left(\frac{u_{i}}{\mu_{u}}\right), &  & \text{\ensuremath{\alpha=}1,}
\end{array}\right.\label{eq:GE}
\end{equation}
where the cases $\alpha=0$ and $\alpha=1$ give the two well-known
inequality measures, the mean logarithmic deviation measure and the
Theil index, respectively. 

Given (\ref{eq:equal_destination}), the extensions to the Class-1
mobility measures considered in section \ref{subsec:Extending-Class-1-mobility}
become the intermediate measures of inequality {\small{}\cite{BoPf:90,Eich:88}:}
\begin{equation}
\frac{\theta\left(c\right)}{n}\sum_{i=1}^{n}\left[\left[\frac{u_{i}+c}{\mu_{u}+c}\right]^{\alpha(c)}-1\right],\:\theta\left(c\right)=\frac{1+c^{2}}{\tilde{\alpha}(c)^{2}-\tilde{\alpha}(c)},\:\alpha(c)\neq0,1;\label{eq:mobility-ext-1}
\end{equation}
in the limit one obtains translation-independent inequality indices.
This limit can be expressed using the class-1 translation-independent
mobility indices $\text{\ensuremath{\mathcal{T}_{\alpha}^{1}}}$ defined
in (\ref{eq:mobility-c1-trans}). If (\ref{eq:equal_destination})
holds then $\text{\ensuremath{\mathcal{T}_{\alpha}^{1}}}$ becomes
\begin{equation}
\left\{ \begin{array}{ccc}
\frac{1}{n\alpha^{2}}\sum_{i=1}^{n}\left[\mathrm{e}^{\alpha\left[u_{i}-\mu_{u}\right]}-1\right], &  & \text{\ensuremath{\alpha\neq}0,}\\
\\
\frac{1}{2}\text{\textrm{var}}\left(u_{i}\right), &  & \text{\ensuremath{\alpha=}0,}
\end{array}\right.\label{eq:Kolm}
\end{equation}
\cite{BlDo:80,BoCo:10}. Members of this class for which $\alpha>0$
are ordinally equivalent to the Kolm class of inequality measures
\cite{Kolm:76I,Cowe:16IA}. 

This link between the mobility index and the inequality index is straightforward
in the case where one has cardinal data; in the case where the data
are ordinal it is necessary to replace the mean by some reference
point -- see \citeN{CoFl:17IW}.

\pagebreak 
\begin{table}[H]
{\small{}}%
\begin{tabular}{l@{}ccccccccccc}
 &  &  &  &  &  &  &  &  &  &  & \tabularnewline
 &  & \multicolumn{2}{c}{{\small{}movement}} &  & \multicolumn{2}{c}{{\small{}consistency}} &  & \multicolumn{2}{c}{{\small{}decomposition}} & {\small{}directional} & \tabularnewline
\hline 
 &  & {\small{}axiom 2} & {\small{}axiom 2'} &  & {\small{}scale} & {\small{}translation} &  & {\small{}up/down} & {\small{}exchange} &  & \tabularnewline
\hline 
\multicolumn{10}{l}{\textsf{\small{}Our indices}} &  & \tabularnewline
{\small{}$\mathcal{A}_{\alpha}^{1}$ in (\ref{eq:mobility-c1-abs})} &  & {\small{}\checkmark} &  &  & {\footnotesize{}(PSI)} &  &  & {\small{}\checkmark} &  & {\small{}\checkmark} & \tabularnewline
{\small{}$\mathcal{A}_{\gamma}^{2}$ in (\ref{eq:mobility-c2-abs})} &  & {\small{}\checkmark} &  &  &  & {\footnotesize{}(PTI)} &  & {\small{}\checkmark} & {\small{}\checkmark} & {\small{}\checkmark} & \tabularnewline
{\small{}$\mathcal{S}_{\alpha}^{1}$ in (\ref{eq:mobility-c1-scale})} &  &  & {\small{}\checkmark} &  & {\small{}\checkmark} &  &  & {\small{}\checkmark} &  & {\small{}\checkmark} & \tabularnewline
{\small{}$\mathcal{S}_{\gamma}^{2}$ in (\ref{eq:mobility-c2-scale})} &  &  & {\small{}\checkmark} &  & {\small{}\checkmark} &  &  & {\small{}\checkmark} & {\small{}\checkmark} & {\small{}\checkmark} & \tabularnewline
{\small{}$\mathcal{T}_{\alpha}^{1}$ in (\ref{eq:mobility-c1-trans})} &  &  & {\small{}\checkmark} &  &  & {\small{}\checkmark} &  & {\small{}\checkmark} &  & {\small{}\checkmark} & \tabularnewline
{\small{}$\mathcal{T}_{\gamma}^{2}$ in (\ref{eq:mobility-c2-trans})} &  &  & {\small{}\checkmark} &  &  & {\small{}\checkmark} &  & {\small{}\checkmark} & {\small{}\checkmark} & {\small{}\checkmark} & \tabularnewline
\hline 
\multicolumn{10}{l}{\textsf{\small{}Other indices}} &  & \tabularnewline
{\small{}$1-\hat{\beta}$ in }(\ref{eq:beta^}) &  &  &  &  & {\footnotesize{}(PSI)} &  &  &  &  &  & \tabularnewline
{\small{}$1-\hat{\rho}$ in }(\ref{eq:corr-coeff}) &  &  &  &  & {\small{}\checkmark} & {\small{}\checkmark} &  &  &  &  & \tabularnewline
{\small{}$FO_{1}$ in (}\ref{eq:FO1}{\small{})} &  & {\small{}\checkmark} &  &  &  & {\footnotesize{}(PTI)} &  & {\small{}\checkmark} & {\small{}\checkmark} &  & \tabularnewline
{\small{}$FO_{2}$ in (}\ref{eq:FO2} {\small{})} &  & {\small{}\checkmark} &  &  & {\footnotesize{}(PSI)} &  &  & {\small{}\checkmark} & {\small{}\checkmark} &  & \tabularnewline
{\small{}$S_{\text{Theil}}$ in (}\ref{eq:Shorrocks}{\small{})} &  & {\small{}\checkmark} &  &  & {\footnotesize{}(PSI)} &  &  &  &  &  & \tabularnewline
{\small{}$S_{\text{Gini}}$ in (}\ref{eq:Shorrocks}{\small{})} &  & {\small{}\checkmark} &  &  & {\footnotesize{}(PSI)} &  &  &  &  &  & \tabularnewline
{\small{}$RG_{1}$ in (}\ref{eq:RG1}{\small{})} &  &  &  &  & {\footnotesize{}(PSI)} &  &  &  &  & {\small{}\checkmark} & \tabularnewline
{\small{}$RG_{2}$ in (}\ref{eq:RG2}{\small{})} &  &  &  &  & {\small{}\checkmark} &  &  &  &  & {\small{}\checkmark} & \tabularnewline
{\small{}$BC_{D}$ in (}\ref{eq:BC_D}{\small{})} &  & {\small{}\checkmark} &  &  & {\footnotesize{}(PSI)} &  &  & {\small{}\checkmark} &  & {\small{}\checkmark} & \tabularnewline
{\small{}$BC_{U}$ in (}\ref{eq:BC_U}{\small{})} &  & {\small{}\checkmark} &  &  & {\footnotesize{}(PSI)} &  &  & {\small{}\checkmark} &  & {\small{}\checkmark} & \tabularnewline
\hline 
\end{tabular}{\small{}\caption{Mobility measures: properties.}
\label{tab:principles}}
\end{table}

\begin{table}[H]
\begin{tabular}{lcccll}
\hline &  &  &  &  & \tabularnewline
Mobility &  &  & \multicolumn{2}{c}{} & Inequality\tabularnewline
\cline{1-2} \cline{2-2} \cline{5-6} \cline{6-6} 
index & parameter &  &  & index & when status $(u_{i},v_{i})=(u_{i},\mu_{u})$\tabularnewline
\cline{1-2} \cline{2-2} \cline{5-6} \cline{6-6} 
 &  &  &  &  & \tabularnewline
Class-1 indices &  &  &  &  & \tabularnewline
$\mathcal{S}_{\alpha}^{1}$ in (\ref{eq:mobility-c1-scale}) & $\alpha=0$ &  &  & MLD & Mean Logarithmic Deviation index\tabularnewline
$\mathcal{S}_{\alpha}^{1}$ in (\ref{eq:mobility-c1-scale}) & $\alpha=1$ &  &  & Theil & Theil index\tabularnewline
$\mathcal{S}_{\alpha}^{1}$ in (\ref{eq:mobility-c1-scale}) & $\alpha\in{\rm I\!R}$ &  &  & GE & Generalised Entropy class of inequality measures\tabularnewline
$\mathcal{T}_{\alpha}^{1}$ in (\ref{eq:mobility-c1-trans}) & $\alpha=0$ &  &  & Var/2 & half the variance\tabularnewline
$\mathcal{T}_{\alpha}^{1}$ in (\ref{eq:mobility-c1-trans}) & $\alpha>0$ &  &  & Kolm & Kolm class of inequality measures\tabularnewline
\cline{1-2} \cline{2-2} \cline{5-6} \cline{6-6} 
 &  &  &  &  & \tabularnewline
Class-2 indices &  &  &  &  & \tabularnewline
$\mathcal{S}_{\gamma}^{2}$ in (\ref{eq:mobility-c2-scale}) & $\gamma=1$ &  &  & $G/(2\mu_{u})$ & half the (relative) Gini index\tabularnewline
$\mathcal{S}_{\gamma}^{2}$ in (\ref{eq:mobility-c2-scale}) & $\gamma>0,\text{odd}$ &  &  & $G_{\gamma}/(2\mu_{u})$ & half the Generalised (relative) Gini index\tabularnewline
$\mathcal{T}_{\gamma}^{2}$ in (\ref{eq:mobility-c2-trans}) & $\gamma=1$ &  &  & $G/2$ & half the absolute Gini index\tabularnewline
$\mathcal{T}_{\gamma}^{2}$ in (\ref{eq:mobility-c2-trans}) & $\gamma>0,\text{odd}$ &  &  & $G_{\gamma}/2$ & half the Generalised absolute Gini index\tabularnewline
$FO_{1}$ in{\small{} (}\ref{eq:FO1}{\small{})} &  &  &  & MAD & mean absolute deviation index\tabularnewline
$FO_{2}$ in{\small{} (}\ref{eq:FO2}{\small{})} &  &  &  & MAD-log & mean absolute log-deviation index\tabularnewline
\hline 
\end{tabular}\caption{Class-1 and Class-2 mobility indices: connections with inequality
measures, when $\forall i:v_{i}=\mu_{u}$.}
\label{tab:connections_inequality}
\end{table}

\paragraph*{Class-2 mobility measures}

We use the class-2 absolute mobility indices $\mathcal{A}_{\gamma}^{2}$,
defined in (\ref{eq:mobility-c2-abs}). Clearly $\mathcal{A}_{0}^{2}=\frac{1}{n}\sum_{i=1}^{n}\left|v_{i}-u_{i}\right|$
-- see (\ref{eq:Mobility-linear}). If (\ref{eq:equal_destination})
holds then this becomes the mean  (absolute) deviation.

From (\ref{eq:relation-Gamma-G}) we have the basic class-2 mobility
measure $\mathcal{A}_{1}^{2}=\nicefrac{1}{2}G+\mu_{d}\left[\frac{1}{2}-p\right]$
where $G$ is the absolute Gini for the status differences $d_{i}$.
But, if we restrict attention to the case where (\ref{eq:equal_destination})
is true, then, for all $i$, $d_{i}=\mu_{u}-u_{i}$ and so $\mu_{d}=0$
. Therefore the mobility measure $\mathcal{A}_{1}^{2}$ becomes simply
half the absolute Gini index of the origin distribution:
\begin{equation}
G(\mathbf{u})=\frac{1}{2n^{2}}\sum_{i=1}^{n}\sum_{j=1}^{n}\mid u_{i}-u_{j}\mid=\frac{1}{n}\left[\sum_{i=1}^{n}\frac{2i}{n}-\frac{1}{n}-1\right]u_{(i)},\label{eq:gini_std}
\end{equation}
More generally for $\gamma=1,3,5,...$ we have $\mathcal{A}_{\gamma}^{2}=\frac{1}{n}\sum_{i=1}^{n}\left[\frac{i}{n}-p-\frac{1}{2n}\right]^{\gamma}d_{(i)}$;
if (\ref{eq:equal_destination}) holds then this becomes%
{} 
\begin{equation}
\frac{1}{n}\sum_{i=1}^{n}c_{i}u_{(i)}\label{eq:Ext_Gini}
\end{equation}
{\small{}where $c_{i}=a_{i}^{\gamma}-\frac{1}{n}\sum_{j=1}^{n}a_{j}^{\gamma}$,
and }$a_{i}${\small{} is given by (}\ref{eq:positional-weights-1}{\small{}).
The expression (}\ref{eq:Ext_Gini}{\small{}) is one version of the
absolute extended Gini for the origin distribution \cite{Chak:88,Weym:81}.
Also note that,} if (\ref{eq:equal_destination}) holds, then $\mathcal{T}_{\gamma}^{2}$
adopts exactly the same form as the limiting form of $\mathcal{A}_{\gamma}^{2}$
-- see equation (\ref{eq:Ext_Gini}); under the same circumstances
$\mathcal{S}_{\gamma}^{2}$ becomes (\ref{eq:Ext_Gini}), divided
by $\mu_{u}$.\footnote{The reason for these two statements is that, given (\ref{eq:equal_destination}),
in the definition (\ref{eq:mobility-c2-trans}) $d_{i}=\left[v_{i}-u_{i}\right]-\left[\mu_{v}-\mu_{u}\right]=\mu_{u}-u_{i}$,
the same as that for the limiting form of (\ref{eq:mobility-c2-abs});
in the definition (\ref{eq:mobility-c2-scale}) $d_{i}=\frac{v_{i}}{\mu_{v}}-\frac{u_{i}}{\mu_{u}}=1-\frac{u_{i}}{\mu_{u}}$,
which is the form of $d_{i}$ adopted by (\ref{eq:mobility-c2-abs})
or (\ref{eq:mobility-c2-trans}), divided by $\mu_{u}$.} 

\medskip{}

We have a unified framework of distributional analysis in which inequality
can be seen as a special case of mobility. For class 1 the mobility
measures $\text{\ensuremath{\mathcal{S}_{\alpha}^{1}}}$ and $\text{\ensuremath{\mathcal{T}_{\alpha}^{1}}}$
are related to the broad class of inequality measures including generalised
entropy, intermediate and Kolm measures; for class 2 measures $\mathcal{A}_{\gamma}^{2}$,
$\mathcal{S}_{\gamma}^{2}$, $\mathcal{T}_{\gamma}^{2}$ the connection
is with the mean deviation (for $\gamma=0$), the Gini index (for
$\gamma=1$), or the generalised Gini (for $\gamma>1$). 

\section{Conclusion and summary\label{sec:conclusion}}

Mobility measurement deserves careful consideration in the way that
the measurement of social welfare, inequality or poverty deserves
careful consideration. This consideration should involve principles,
formal reasoning and empirical applicability. However, the bulk of
empirical studies of mobility analysis apply ready-made techniques
that, in this application, are seriously flawed. The flaws matter
because, in certain circumstances, the ready-made techniques give
exactly the wrong guidance on basic questions such as ``does scenario
A exhibit more movement of persons than scenario B?''

\subsection{An approach to mobility}

Our approach is to show that a consistent theory of mobility measurement
can be founded on three basic principles of mobility comparisons.
We do this using the methodology of \citeANP{CoFl:17IW} (\citeyearNP{CoFl:17IW,CoFl:18MM})
to provide a natural interpretation of these principles in terms of
formal axioms. These axioms are used in theorems that characterise
two classes of mobility measures that can be easily implemented in
terms of income (wealth) mobility or rank mobility. 

\subsection{Overview of mobility measures}

Table \ref{tab:principles} summarises the properties of the mobility
measures used in this paper, with respect to the principles discussed
in section \ref{subsec:Principles}. 
\begin{itemize}
\item The "more movement, more mobility" principle has two versions, defined
by axioms \ref{ax:monotonicity} and 2' on monotonicity (see sections
\ref{subsec:ordering-structure} and \ref{subsec:aggregate-mobility}). 
\item The consistency principle covers the scale-independence and translation-independence
properties (see sections \ref{subsec:ordering-scale} and \ref{subsec:aggregate-mobility}).
Weaker versions are reported in parenthesis when indices are unchanged
only by a scale change (PSI) or by a level change (PTI) in the two
periods for the {same} amount (see sections \ref{subsec:Consistency-concepts}
and \ref{subsec:Translation-consistency}). 
\item The decomposition principle covers the distinction between upward
and downward movements, and the decomposition with respect to structural
mobility and growth.
\end{itemize}
It is interesting to notice that decomposition with respect to exchange
mobility and growth is available with class-2 mobility indices ($\mathcal{A}_{\gamma}^{2}$,
$\mathcal{S}_{\gamma}^{2}$, $\mathcal{T}_{\gamma}^{2},FO_{1},FO_{2}$),
because they are expressed as weighted means of status.\footnote{Remember that Fields-Ok indices $FO_{1}$ and $FO_{2}$ are special
cases of class-2 mobility indices, see section \ref{subsec:Non-normalised-status}.}

\subsection{Mobility and inequality}

Finally, as discussed in section \ref{subsec:Mobility-and-inequality},
there are close connections between class-1 and class-2 mobility measures
and several well-known inequality measures. Mobility indices are reduced
to inequality measures when the destination is an equal distribution
with the same mean as the origin distribution ($\forall i:v_{i}=\mu_{u}$).
Table \ref{tab:connections_inequality} summarises these relationships.
We can see that: 
\begin{itemize}
\item Class-1 indices reduce to the Generalised Entropy class of inequality
measures, with the Theil and MLD indices as special cases, for scale-independent
measures, and to half the variance or the Kolm indices for translation-independent
indices. 
\item Class-2 indices reduce to half the (generalised) relative Gini index
for scale-independent measures, half the (generalised) absolute Gini
index for translation-independent measures, and to mean absolute deviation
indices for Fields-Ok measures. 
\end{itemize}
Overall, class-1 and class-2 mobility measures share the same theoretical
foundations as inequality measures.

\vfill{}

\pagebreak{}

\bibliographystyle{chicago}
\bibliography{combine}

\pagebreak

\appendix
{}

\section*{Appendix: Proofs}

\subsection*{\uline{Theorem \mbox{\ref{th:additive-structure}} }}

\begin{proof} In both the case where $Z$ is a connected subset of
$\mathbb{R}\times\mathbb{R}$ and the case where $Z$ is $\mathbb{Q}_{+}\times\mathbb{Q}_{+}$
Theorem 5.3 of \citeN{Fish:70} can be invoked to show that axioms
\ref{ax:continuity} to \ref{ax:independence} imply that $\succeq$
can be represented as 
\begin{equation}
\sum_{i=1}^{n}\phi_{i}\left(z_{i}\right),\forall\mathbf{z}\in Z^{n},\label{eq:basic-linear-1}
\end{equation}
where $\phi_{i}$ is continuous, defined up to an affine transformation
and, by Axiom \ref{ax:monotonicity}, is increasing in $v_{i}$ if
$v_{i}>u_{i}$ and \emph{vice versa}. Using Axiom \ref{ax:local-immobility}
in (\ref{eq:basic-linear-1}) we have
\begin{equation}
\phi_{i}\left(u_{i},u_{i}\right)=\phi_{i}\left(u_{i}+\delta,u_{i}+\delta\right),\label{eq:basic-linear-with-immob}
\end{equation}
where $\delta:=u_{i}'-u_{i}$. Equation (\ref{eq:basic-linear-with-immob})
implies that $\phi_{i}$ must take the form $\phi_{i}\left(u,u\right)=a_{i}+b_{i}u$.
Since $\phi_{i}$ is defined up to an affine transformation we may
choose $a_{i}=0$ and so we have
\begin{equation}
\phi_{i}\left(u,u\right)=b_{i}u.\label{eq:phi_i(x,x)}
\end{equation}

\end{proof}

\bigskip{}

\subsection*{\uline{Theorem \mbox{\ref{th:Relation-between-axioms}}}}

\begin{proof} In parts (i) to (iii) it is assumed that (\ref{eq:equivalent_profiles})
holds and that $\lambda,\mu$ are positive constants.

\paragraph*{(i) OSI and DSI imply PSI.}

OSI (\ref{eq:OSI}) means $\left(\lambda\mathbf{u,v}\right)\sim\left(\lambda\mathbf{u^{\prime},v^{\prime}}\right)$.
If DSI (\ref{eq:DSI}) also holds then 
\begin{equation}
\left(\lambda\mathbf{u,\mu v}\right)\sim\left(\lambda\mathbf{u^{\prime},\mu v^{\prime}}\right).\label{eq:OSI+DSI-implication}
\end{equation}
Putting $\mu=\lambda$ in (\ref{eq:OSI+DSI-implication}) we obtain
PSI (\ref{eq:PSI}). 

\paragraph*{(ii) OSI and PSI imply DSI.}

OSI (\ref{eq:OSI}) means $\left(\mu\mathbf{u,v}\right)\sim\left(\mu\mathbf{u^{\prime},v^{\prime}}\right).$
If PSI (\ref{eq:PSI}) also holds then 
\begin{equation}
\left(\lambda\mu\mathbf{u,\lambda v}\right)\sim\left(\lambda\mathbf{\mu u^{\prime},\lambda v^{\prime}}\right).\label{eq:OSI+PSI-implication}
\end{equation}
Putting $\mu=1/\lambda$ in (\ref{eq:OSI+PSI-implication}) we obtain
DSI (\ref{eq:DSI}).

\paragraph*{(iii) DSI and PSI imply OSI. }

DSI (\ref{eq:DSI}) means $\left(\mathbf{u,\mu v}\right)\sim\left(\mathbf{u^{\prime},\mu v^{\prime}}\right).$
If PSI (\ref{eq:PSI}) also holds then 
\begin{equation}
\left(\lambda\mathbf{u,\lambda\mu v}\right)\sim\left(\lambda\mathbf{u^{\prime},\lambda\mu v^{\prime}}\right).\label{eq:DSI+PSI-implication}
\end{equation}
Putting $\mu=1/\lambda$ in (\ref{eq:DSI+PSI-implication}) we obtain
OSI (\ref{eq:OSI}).

\end{proof}

\bigskip{}

\subsection*{\uline{Theorem \mbox{\ref{th:scale-three-defs}}}}

\begin{proof} In parts (i) to (iii) it is assumed that (\ref{eq:equivalent_profiles})
holds and that $\lambda,\mu$ are positive constants. 

\paragraph*{(i) Case OSI. }

Theorem\textbf{ \ref{th:additive-structure}} and OSI (\ref{eq:OSI})
together imply 
\[
\sum_{i=1}^{n}\phi_{i}\left(u_{i},v_{i}\right)=\sum_{i=1}^{n}\phi_{i}\left(u'_{i},v'_{i}\right),\sum_{i=1}^{n}\phi_{i}\left(\lambda u_{i},v_{i}\right)=\sum_{i=1}^{n}\phi_{i}\left(\lambda u'_{i},v'_{i}\right)
\]
 and so:
\[
\frac{\sum_{i=1}^{n}\phi_{i}\left(u'_{i},v'_{i}\right)}{\sum_{i=1}^{n}\phi_{i}\left(u_{i},v_{i}\right)}=\frac{\sum_{i=1}^{n}\phi_{i}\left(\lambda u'_{i},v'_{i}\right)}{\sum_{i=1}^{n}\phi_{i}\left(\lambda u_{i},v{}_{i}\right)}.
\]
Therefore the function (\ref{eq:basic-linear}) is homothetic in $u$
so that we may write

\begin{equation}
\sum_{i=1}^{n}\phi_{i}\left(\lambda u_{i},v{}_{i}\right)=\theta\left(\lambda,\sum_{i=1}^{n}\phi_{i}\left(u_{i},v_{i}\right)\right),\label{eq:homothetic-additive_OSI}
\end{equation}
where $\theta:\mathbb{R\rightarrow R}$ is increasing in its second
argument. Consider the case where, for arbitrary distinct values $j$
and $k$, we have $\phi_{i}\left(u_{i},v_{i}\right)=0$ for all $i\neq j,k$;
then, for given values of $v_{j},v_{k},\lambda$, (\ref{eq:homothetic-additive_OSI})
can be written as the functional equation:

\begin{equation}
f_{j}\left(u_{j}\right)+f{}_{k}\left(u_{k}\right)=h\left(g_{j}\left(u_{j}\right)+g{}_{k}\left(u_{k}\right)\right),\label{eq:functional-additive_OSI}
\end{equation}
where $f_{i}\left(u\right):=\phi_{i}\left(\lambda u,v_{i}\right),$
$g_{i}\left(u\right):=\phi_{i}\left(u,v_{i}\right),\;i=j,k$ and $h\left(x\right):=\theta\left(\lambda,x\right)$.
Equation (\ref{eq:functional-additive_OSI}) has the solution 
\begin{eqnarray*}
f_{i}\left(u\right) & = & a_{0}g{}_{i}\left(u\right)+a{}_{i},\;i=j,k,\\
h\left(x\right) & = & a_{0}x+a{}_{j}+a{}_{k},
\end{eqnarray*}
where $a_{0},a_{j},a_{k}$, are constants that may depend on $\lambda,v_{j},v_{k}$
(\citeNP{PoZa:04}, Supplement S.5.5). Therefore:
\begin{eqnarray}
\phi_{j}\left(\lambda u{}_{j},v_{j}\right) & = & a_{0}\left(\lambda,v_{j},v_{k}\right)\phi_{j}\left(u_{j},v_{j}\right)+a{}_{j}\left(\lambda,v_{j},v_{k}\right)\label{eq:phi_j_OSI}\\
\phi_{k}\left(\lambda u{}_{k},v_{k}\right) & = & a_{0}\left(\lambda,v_{j},v_{k}\right)\phi_{k}\left(u_{k},v_{k}\right)+a{}_{k}\left(\lambda,v_{j},v_{k}\right).\label{eq:phi_k_OSI}
\end{eqnarray}
Since $j$ and $k$ are arbitrary, we could repeat the analysis for
arbitrary distinct values $j$ and $\ell$ and $v_{i}=u{}_{i}=0$
for all $i\neq j,\ell$, where $\ell\neq k$; then we would have
\begin{eqnarray}
\phi_{j}\left(\lambda u{}_{j},v_{j}\right) & = & a'_{0}\left(\lambda,v_{j},v_{k}\right)\phi_{j}\left(u_{j},v_{j}\right)+a'_{j}\left(\lambda,v_{j},v_{\ell}\right)\label{eq:phi_j-1_OSI}\\
\phi_{\ell}\left(\lambda u_{\ell},v_{\ell}\right) & = & a'_{0}\left(\lambda,v_{j},v_{\ell}\right)\phi_{\ell}\left(u_{\ell},v_{\ell}\right)+a'_{\ell}\left(\lambda,v_{j},v_{\ell}\right).\label{eq:phi_ell_OSI}
\end{eqnarray}
where $a'_{0},a'_{j},a_{\ell}$, are constants that may depend on
$\lambda,v_{j},v_{\ell}$. The right-hand sides of (\ref{eq:phi_j_OSI})
and (\ref{eq:phi_j-1_OSI}) are equal and so $a_{j}$ must be independent
of $v_{k}$ and $a_{0}$ must be independent of $v_{j}$, $v_{k}$.
Therefore, because $j$ and $k$ are arbitrary, for all $i$ and for
any values of $u,v$: 
\begin{equation}
\phi_{i}\left(\lambda u,v\right)=a_{0}\left(\lambda\right)\phi_{i}\left(u,v\right)+a{}_{i}\left(\lambda,v\right).\label{eq:phi_i_OSI-1}
\end{equation}
Taking the special case where $u=0$ we get $\phi_{i}\left(0,v\right)=a_{0}\left(\lambda\right)\phi_{i}\left(0,v\right)+a{}_{i}\left(\lambda,v\right)$
and so, defining $\psi_{i}\left(u,v\right):=\phi_{i}\left(u,v\right)-\phi_{i}\left(0,v\right)$,
equation (\ref{eq:phi_i_OSI-1}) becomes 
\begin{equation}
\psi_{i}\left(\lambda u,v\right)=a_{0}\left(\lambda\right)\psi_{i}\left(u,v\right).\label{eq:phi_i_OSI-2}
\end{equation}
Equation (\ref{eq:phi_i_OSI-2}) can be expressed as $f\left(x+y\right)=g(y)+f(x)$
where $x=\log u_{i}$, $y=\log\lambda$, $f\left(x\right):=\log\psi_{i}\left(u_{i},v_{i}\right)$,
$g\left(y\right):=\log a_{0}\left(\lambda\right)$. This Pexider equation
has the solution $f\left(x\right)$=$\alpha x+\beta,$ $g\left(y\right)=\alpha y$
which implies 
\begin{equation}
\log\psi_{i}\left(u,v\right)=\alpha\log u_{i}+\log A,\;\;\;\log a_{0}\left(\lambda\right)=\alpha\log\lambda\label{eq:psi_i_solution_OSI}
\end{equation}
Hence, for a given specific $v_{i}$ , from the definition of $\psi_{i}$
we have $\log\left(\phi_{i}\left(u,v\right)-\phi_{i}\left(0,v\right)\right)=\alpha\log u+\log A_{i}\left(v\right)$
and so, for any $u,v$: 
\begin{equation}
\phi_{i}\left(u,v\right)=\phi_{i}\left(0,v\right)+u^{\alpha}A_{i}\left(v\right).\label{eq:phi_i_result_OSI}
\end{equation}
In the specific case where $u=v$, (\ref{eq:phi_i(u,u)}) implies
that (\ref{eq:phi_i_result_OSI}) becomes $0=\phi_{i}\left(0,v\right)+A_{i}\left(v\right)v^{\alpha},$
which means that (\ref{eq:phi_i_result_OSI}) can be written as 
\begin{equation}
\phi_{i}\left(u,v\right)=A_{i}\left(v\right)\left[u^{\alpha}-v^{\alpha}\right].\label{eq:phi_i_solution_OSI}
\end{equation}

\paragraph*{(ii) Case DSI. }

Theorem\textbf{ \ref{th:additive-structure}} and OSI (\ref{eq:DSI})
together imply 
\[
\sum_{i=1}^{n}\phi_{i}\left(u_{i},v_{i}\right)=\sum_{i=1}^{n}\phi_{i}\left(u'_{i},v'_{i}\right),\sum_{i=1}^{n}\phi_{i}\left(u_{i},\lambda v_{i}\right)=\sum_{i=1}^{n}\phi_{i}\left(u'_{i},\lambda v'_{i}\right).
\]
 Using the same argument as in case OSI the function (\ref{eq:basic-linear})
may be written

\begin{equation}
\sum_{i=1}^{n}\phi_{i}\left(u_{i},v{}_{i}\right)=\theta\left(\lambda,\sum_{i=1}^{n}\phi_{i}\left(u_{i},\lambda v_{i}\right)\right),\label{eq:homothetic-additive-DSI}
\end{equation}
where $\theta:\mathbb{R\rightarrow R}$ is increasing in its second
argument. The rest of the proof follows immediately from the proof
of case OSI, interchanging the roles of $u$ and $v$ and giving
\begin{equation}
\phi_{i}\left(u,v\right)=A'_{i}\left(u\right)\left[v^{\alpha}-u^{\alpha}\right].\label{eq:phi_i_solution_DSI}
\end{equation}

\paragraph*{(iii) Case PSI. }

Theorem\textbf{ \ref{th:additive-structure}} and OSI (\ref{eq:PSI})
together imply $\sum_{i=1}^{n}\phi_{i}\left(\lambda z_{i}\right)=\sum_{i=1}^{n}\phi_{i}\left(\lambda z_{i}^{\prime}\right)$
which further implies 
\[
\frac{\sum_{i=1}^{n}\phi_{i}\left(z_{i}^{\prime}\right)}{\sum_{i=1}^{n}\phi_{i}\left(z_{i}\right)}=\frac{\sum_{i=1}^{n}\phi_{i}\left(\lambda z_{i}^{\prime}\right)}{\sum_{i=1}^{n}\phi_{i}\left(\lambda z_{i}\right)}
\]
so that the function (\ref{eq:basic-linear}) is homothetic and we
have:

\begin{equation}
\sum_{i=1}^{n}\phi_{i}\left(\lambda z_{i}\right)=\theta\left(\lambda,\sum_{i=1}^{n}\phi_{i}\left(z_{i}\right)\right),\label{eq:homothetic-additive_PSI}
\end{equation}
where $\theta:\mathbb{R\rightarrow R}$ is increasing in its second
argument. Consider the case where, for arbitrary distinct values $j$
and $k$, we have$\phi_{i}\left(u_{i},v_{i}\right)=0$ for all $i\neq j,k$;
then, for given values of $v_{j},v_{k},\lambda$, (\ref{eq:homothetic-additive_PSI})
can be written as the functional equation: 
\begin{equation}
f_{j}\left(u_{j}\right)+f{}_{k}\left(u_{k}\right)=h\left(g_{j}\left(u_{j}\right)+g{}_{k}\left(u_{k}\right)\right),\label{eq:functional-additive_PSI}
\end{equation}
where $f_{i}\left(u\right):=\phi_{i}\left(\lambda u,\lambda v_{i}\right),$
$g_{i}\left(u\right):=\phi_{i}\left(u,v_{i}\right),\;i=j,k$ and $h\left(x\right):=\theta\left(\lambda,x\right)$.
Equation (\ref{eq:functional-additive_PSI}) has the solution 
\begin{eqnarray*}
f_{i}\left(u\right) & = & a_{0}g{}_{i}\left(u\right)+a{}_{i},\;i=j,k,\\
h\left(x\right) & = & a_{0}x+a{}_{j}+a{}_{k},
\end{eqnarray*}
where $a_{0},a_{j},a_{k}$, are constants that may depend on $\lambda,v_{j},v_{k}$
(\citeNP{PoZa:04}, Supplement S.5.5). Therefore:
\begin{eqnarray}
\phi_{j}\left(\lambda u{}_{j},\lambda v_{j}\right) & = & a_{0}\left(\lambda,v_{j},v_{k}\right)\phi_{j}\left(u_{j},v_{j}\right)+a{}_{j}\left(\lambda,v_{j},v_{k}\right)\label{eq:phi_j_PSI}\\
\phi_{k}\left(\lambda u{}_{k},\lambda v_{k}\right) & = & a_{0}\left(\lambda,v_{j},v_{k}\right)\phi_{k}\left(u_{k},v_{k}\right)+a{}_{k}\left(\lambda,v_{j},v_{k}\right).\label{eq:phi_k_PSI}
\end{eqnarray}
Since $j$ and $k$ are arbitrary, we could repeat the analysis for
arbitrary distinct values $j$ and $\ell$ and $v_{i}=u{}_{i}=0$
for all $i\neq j,\ell$, where $\ell\neq k$; then we would have
\begin{eqnarray}
\phi_{j}\left(\lambda u{}_{j},\lambda v_{j}\right) & = & a'_{0}\left(\lambda,v_{j},v_{k}\right)\phi_{j}\left(u_{j},v_{j}\right)+a'_{j}\left(\lambda,v_{j},v_{\ell}\right)\label{eq:phi_j-1_PSI}\\
\phi_{\ell}\left(\lambda u_{\ell},\lambda v_{\ell}\right) & = & a'_{0}\left(\lambda,v_{j},v_{\ell}\right)\phi_{\ell}\left(u_{\ell},v_{\ell}\right)+a'_{\ell}\left(\lambda,v_{j},v_{\ell}\right).\label{eq:phi_ell_PSI}
\end{eqnarray}
where $a'_{0},a'_{j},a_{\ell}$, are constants that may depend on
$\lambda,v_{j},v_{\ell}$. The right-hand sides of (\ref{eq:phi_j_PSI})
and (\ref{eq:phi_j-1_PSI}) are equal and so $a_{j}$ must be independent
of $v_{j}$ and $a_{0}$ must be independent of $v_{j}$, $v_{k}$.
Therefore, because $j$ and $k$ are arbitrary, for all $i$ we have
\begin{equation}
\phi_{i}\left(\lambda u{}_{i},\lambda v_{i}\right)=a_{0}\left(\lambda\right)\phi_{i}\left(u_{i},v_{i}\right)+a{}_{i}\left(\lambda,v_{i}\right).\label{eq:phi_i_PSI}
\end{equation}
In the case where $v_{i}=u_{i}$, (\ref{eq:phi_i(u,u)}) and (\ref{eq:phi_i_PSI})
yield
\[
a_{i}\left(\lambda,v_{i}\right)=0
\]
so that (\ref{eq:phi_i_PSI}) can be rewritten

\begin{equation}
\phi_{i}\left(\lambda u{}_{i},\lambda v_{i}\right)=a_{0}\left(\lambda\right)\phi_{i}\left(u_{i},v_{i}\right).\label{eq:phi_i-1_PSI}
\end{equation}
 From \citeN{AcDh:89}, page 346 there must exist $\beta\in\mathbb{R}$
and a function $h:\mathbb{R}_{+}\rightarrow\mathbb{R}$ \ such that
$\phi'_{i}\left(u,v\right)=v^{\beta}h_{i}\left(u/v\right)$ , so that
\begin{equation}
\phi{}_{i}\left(u,v\right)=v^{\beta}h_{i}\left(\frac{u}{v}\right).\label{eq:phi_i_solution_PSI-1-2}
\end{equation}
From (\ref{eq:phi_i(u,u)}) we see that (\ref{eq:phi_i_solution_PSI-1-2})
implies $h_{i}\left(1\right)=0$. 

\end{proof}

\bigskip{}

\subsection*{\uline{Translation invariance}}

\begin{proof} In parts (i) to (iii) it is assumed that (\ref{eq:equivalent_profiles})
holds and that $\delta,\epsilon$ are constants.

\paragraph*{(i) OTI and DTI imply PTI.}

OTI means $\left(\mathbf{u+\delta1,v}\right)\sim\left(\mathbf{u^{\prime}+\delta1,v^{\prime}}\right)$.
If DTI (\ref{eq:DTI}) also holds then 
\begin{equation}
\left(\mathbf{u+\delta1,v+\epsilon1}\right)\sim\left(\mathbf{u^{\prime}+\delta1,v^{\prime}+\epsilon1}\right).\label{eq:OTI+DTI-implication}
\end{equation}
Putting $\epsilon=\delta$ in (\ref{eq:OTI+DTI-implication}) we obtain
PTI ( \ref{eq:PTI}). 

\paragraph*{(ii) OTI and PTI imply DTI.}

OTI means $\left(\mathbf{u+\delta1,v}\right)\sim\left(\mathbf{u^{\prime}+\delta1,v^{\prime}}\right)$
If PTI also holds then 
\begin{equation}
\left(\mathbf{u+\delta1+\epsilon1,v+\epsilon1}\right)\sim\left(\mathbf{u^{\prime}+\delta1+\epsilon1,v^{\prime}+\epsilon1}\right).\label{eq:OTI+PTI-implication}
\end{equation}
Putting $\epsilon=-\delta$ in (\ref{eq:OTI+PTI-implication}) we
obtain DTI (\ref{eq:DTI}).

\paragraph*{(iii) DTI and PTI imply OTI. }

DTI (\ref{eq:DTI}) means $\left(\mathbf{u,v+\delta1}\right)\sim\left(\mathbf{u^{\prime},v^{\prime}+\delta1}\right)$.
If PTI also holds then 
\begin{equation}
\left(\mathbf{u+\epsilon1,v+\delta1+\epsilon1}\right)\sim\left(\mathbf{u^{\prime}+\epsilon1,v^{\prime}+\delta1+\epsilon1}\right).\label{eq:DTI+PTI-implication}
\end{equation}
Putting $\epsilon=-\delta$ in (\ref{eq:DTI+PTI-implication}) we
obtain OTI (\ref{eq:OTI}).

\end{proof}

\bigskip{}

\subsection*{\uline{Theorem \mbox{\ref{th:PTI}}}}

\begin{proof} 

\paragraph*{(i) Case OTI. }

Theorem\textbf{ \ref{th:additive-structure}}, equation \textbf{(}\ref{eq:equivalent_profiles})
and OTI imply $\sum_{i=1}^{n}\phi_{i}\left(u_{i},v_{i}\right)= \sum_{i=1}^{n}\phi_{i}\left(u'_{i},v_{i}\right)$, $\sum_{i=1}^{n}\phi_{i}\left(\lambda u_{i},v_{i}\right)= \sum_{i=1}^{n}\phi_{i}\left(\lambda u'_{i},v_{i}\right),$which
implies:
\begin{eqnarray}
\sum_{i=1}^{n}\phi_{i}\left(u_{i}+\delta,v_{i}\right) & = & \sum_{i=1}^{n}\phi_{i}\left(u'_{i}+\delta,v'_{i}\right).\label{eq:phi_translation_OTI}
\end{eqnarray}
For given values of $v_{i}$ write 
\begin{equation}
\psi_{i}\left(u\right):=\phi_{i}\left(u,v_{i}\right).\label{eq:phi-to-psi}
\end{equation}
Then (\ref{eq:phi_translation_OTI}) implies:
\[
\frac{\sum_{i=1}^{n}\psi_{i}\left(u'_{i}\right)}{\sum_{i=1}^{n}\psi_{i}\left(u_{i}\right)}=\frac{\sum_{i=1}^{n}\psi_{i}\left(\delta+u'_{i}\right)}{\sum_{i=1}^{n}\psi_{i}\left(\delta+u_{i}\right)}.
\]
Therefore the function (\ref{eq:basic-linear}) has the property that 

\begin{equation}
\sum_{i=1}^{n}\psi_{i}\left(\delta+u_{i}\right)=\theta\left(\delta,\sum_{i=1}^{n}\psi_{i}\left(u_{i}\right)\right),\label{eq:homothetic-additive_OTI}
\end{equation}
 Consider the case where, for arbitrary distinct values $j$ and $k$,
we have $\psi_{i}\left(u_{i}\right)=0$ for all $i\neq j,k$; then,
for given values of $v_{j},v_{k},\lambda$, (\ref{eq:homothetic-additive_PSI})
can be written as the functional equation: 
\begin{equation}
f_{j}\left(u_{j}\right)+f{}_{k}\left(u_{k}\right)=h\left(g_{j}\left(u_{j}\right)+g{}_{k}\left(u_{k}\right)\right),\label{eq:functional-additive_OTI_0}
\end{equation}
where $f_{i}\left(u\right):=\psi_{i}\left(\delta+u\right),$ $g_{i}\left(u\right):=\psi_{i}\left(u\right),\;i=j,k$
and $h\left(x\right):=\theta\left(\delta,x\right)$. Equation (\ref{eq:functional-additive_OTI_0})
has the solution 
\begin{eqnarray*}
f_{i}\left(u\right) & = & a_{0}g{}_{i}\left(u\right)+a{}_{i},\;i=j,k,\\
h\left(x\right) & = & a_{0}x+a{}_{j}+a{}_{k},
\end{eqnarray*}
where $a_{0},a_{j},a_{k}$, are constants that may depend on $\lambda,v_{j},v_{k}$
(\citeNP{PoZa:04}, Supplement S.5.5). Therefore: 
\begin{eqnarray}
\psi_{j}\left(\delta+u_{j}\right) & = & a_{0}\left(\delta\right)\psi_{j}\left(u_{j}\right)+a{}_{j}\left(\delta\right)\label{eq:psi_j_OTI}\\
\psi_{k}\left(\delta+u_{k}\right) & = & a_{0}\left(\delta\right)\psi_{k}\left(u_{k}\right)+a{}_{k}\left(\delta\right).\label{eq:psi_k_OTI}
\end{eqnarray}
Because $j$ and $k$ are arbitrary, for all $i$ we have 
\begin{equation}
\psi_{i}\left(\delta+u\right)=a_{0}\left(\delta\right)\psi_{i}\left(u\right)+a{}_{i}\left(\delta\right).\label{eq:psi_i_OTI}
\end{equation}
Equation (\ref{eq:psi_i_OTI}) is a generalised Pexider equation with
two solutions:\footnote{See \citeN{AcDh:89} and
https://eqworld.ipmnet.ru/en/solutions/fe/fe4107.pdf} (1) $\psi_{i}\left(u\right)=c_{1}u+c_{2},a_{0}\left(\delta\right)=1,a_{i}\left(\delta\right)=c_{1}\delta$,
and (2) $\psi_{i}\left(u\right)=c_{1}e^{\beta u}+c_{2},a_{0}\left(\delta\right)=e^{\beta u},a_{i}\left(\delta\right)=\left[1-e^{\beta u}\right]$.
Substitute back for $\phi_{i}$ using (\ref{eq:phi-to-psi}); then,
using equation (\ref{eq:phi_i(u,u)}) and an appropriate normalisation\footnote{\label{fn:The-normalisation-is}The normalisation is appropriate in
that it ensures that solution 1 is the limiting form of solution 2
as the parameter $\beta\rightarrow0$. To see this, note that the
Taylor expansion of $e^{x}$is $\sum_{n=0}^{\infty}\frac{1}{n!}x^{n}$;
$\frac{1}{\beta}\left[e^{\beta x}-1\right]$ can be written as $\left[1+\beta x+\nicefrac{1}{2}\beta^{2}x^{2}+...-1\right]/\beta$;
the limit of this expression as $\beta\rightarrow0$ is $x.$} we get
\begin{description}
\item [{Solution}] 1 %
$\quad\quad\phi_{i}\left(u,v\right)=A_{i}\left(v\right)\left[u-v\right]$,
\item [{Solution}] 2 %
\qquad{}$\phi_{i}\left(u,v\right)=A_{i}\left(v\right)\frac{1}{\beta}\left[e^{\beta\left[u-v\right]}-1\right]$.
\end{description}
\bigskip{}

\paragraph*{(ii) Case DTI.}

The proof of this case follows immediately from that of Case (i) (OTI),
by interchanging $u$ and $v$. We get
\begin{description}
\item [{Solution}] 1 $\quad\quad\phi_{i}\left(u,v\right)=A'_{i}\left(u\right)\left[v-u\right]$
\item [{Solution}] 2 \qquad{}$\phi_{i}\left(u,v\right)=A'_{i}\left(u\right)\frac{1}{\beta}\left[e^{\beta\left[v-u\right]}-1\right]$
\end{description}
\bigskip{}

\paragraph*{(iii) Case PTI. }

Theorem\textbf{ \ref{th:additive-structure}}, equation \textbf{(}\ref{eq:equivalent_profiles})
and PTI imply 
\begin{eqnarray}
\sum_{i=1}^{n}\phi_{i}\left(u_{i}+\delta,v_{i}+\delta\right) & = & \sum_{i=1}^{n}\phi_{i}\left(u'_{i}+\delta,v'_{i}+\delta\right).\label{eq:phi_translation_PTI-1}
\end{eqnarray}
which means that 

\begin{equation}
\sum_{i=1}^{n}\phi_{i}\left(\delta+u_{i},\delta+v_{i}\right)=\theta\left(\delta,\sum_{i=1}^{n}\phi_{i}\left(u_{i},v_{i}\right)\right),\label{eq:homothetic-additive_PTI}
\end{equation}
where $\theta:\mathbb{R\rightarrow R}$ is increasing in its second
argument. Consider the case where, for arbitrary distinct values $j$
and $k$, we have $\phi_{i}\left(u_{i},v_{i}\right)=0$ for all $i\neq j,k$;
then, for given values of $v_{j},v_{k},\lambda$, (\ref{eq:homothetic-additive_PSI})
can be written as the functional equation: 
\begin{equation}
f_{j}\left(u_{j}\right)+f{}_{k}\left(u_{k}\right)=h\left(g_{j}\left(u_{j}\right)+g{}_{k}\left(u_{k}\right)\right),\label{eq:functional-additive_PTI}
\end{equation}
where $f_{i}\left(u\right):=\phi_{i}\left(\delta+u,\delta+v_{i}\right),$
$g_{i}\left(u\right):=\phi_{i}\left(u,v_{i}\right),\;i=j,k$ and $h\left(x\right):=\theta\left(\delta,x\right)$.
Equation (\ref{eq:functional-additive_OSI}) has the solution 
\begin{eqnarray*}
f_{i}\left(u\right) & = & a_{0}g{}_{i}\left(u\right)+a{}_{i},\;i=j,k,\\
h\left(x\right) & = & a_{0}x+a{}_{j}+a{}_{k},
\end{eqnarray*}
where $a_{0},a_{j},a_{k}$, are constants that may depend on $\delta,v_{j},v_{k}$
(\citeNP{PoZa:04}, Supplement S.5.5). Therefore: 
\begin{eqnarray}
\phi_{j}\left(\delta+u_{j},\delta+v_{j}\right) & = & a_{0}\left(\delta\right)\phi_{j}\left(u_{j},v_{j}\right)+a{}_{j}\left(\delta\right)\label{eq:phi_j_PTI}\\
\phi_{k}\left(\delta+u_{k},\delta+v_{k}\right) & = & a_{0}\left(\delta\right)\phi_{k}\left(u_{k},v_{k}\right)+a{}_{k}\left(\delta\right).\label{eq:phi_k_PTI}
\end{eqnarray}
 Therefore, because $j$ and $k$ are arbitrary, for any $i$ we have
\begin{equation}
\phi_{i}\left(\delta+u,\delta+v\right)=a_{0}\left(\delta\right)\phi_{i}\left(u,v\right)+a{}_{i}\left(\delta\right).\label{eq:phi_i_PTI}
\end{equation}
Using equation (\ref{eq:phi_i(u,u)}) and putting $u=v$ in (\ref{eq:phi_i_PTI})%
{} we have $a_{i}\left(\delta\right)=0,$ from which (\ref{eq:phi_i_PTI})
yields%

\begin{equation}
\phi_{i}\left(\delta+u,\delta+v\right)=a_{0}\left(\delta\right)\phi_{i}\left(u,v\right).\label{eq:phi_i_PTI_1}
\end{equation}
Putting $\delta=-u$ and then $\delta=-v$ in (\ref{eq:phi_i_PTI_1})
we obtain, respectively 
\begin{eqnarray}
\phi_{i}\left(0,v-u\right) & = & a_{0}\left(-u\right)\phi_{i}\left(u,v\right),\label{eq:phi_i_PTI_2}\\
\phi_{i}\left(u-v,0\right) & = & a_{0}\left(-v\right)\phi_{i}\left(u,v\right).\label{eq:phi_i_PTI_3}
\end{eqnarray}
Rearranging (\ref{eq:phi_i_PTI_2}) and (\ref{eq:phi_i_PTI_3}) we
have 
\begin{equation}
\log\left(\phi_{i}\left(0,v-u\right)/\phi_{i}\left(v-u,0\right)\right)=\log a_{0}\left(-u\right)-\log a_{0}\left(-v\right).\label{eq:phi_i_PTI_4}
\end{equation}
Equation (\ref{eq:phi_i_PTI_4}) is a Pexider equation with solution
$\log a_{0}\left(-u\right)=-au$, where $a$ is a constant. Using
this result in (\ref{eq:phi_i_PTI_2}) we find 

\begin{equation}
\phi_{i}\left(u,v\right)=e^{au}g_{i}\left(v-u\right),\label{eq:phi_j_PTI_4}
\end{equation}
where $g_{i}\left(x\right):=\phi_{i}\left(0,x\right)$ and, using
equation (\ref{eq:phi_i(u,u)}), $g_{i}\left(0\right)=0$. Alternatively
we have

\begin{equation}
\phi_{i}\left(u,v\right)=e^{av}g'_{i}\left(u-v\right),\label{eq:phi_j_PTI_5}
\end{equation}
where $g'_{i}\left(x\right):=\phi_{i}\left(0,x\right)$ and $g'_{i}\left(0\right)=0$.
\end{proof}

\subsection*{Structural, exchange, growth decomposition }

\textbf{Absolute mobility index}: using $d_{i}^{\mathsf{A}}=v_{i}-u_{i}$
and $d_{i}^{\mathsf{A'}}=v_{i}-u'_{i}$, we have 
\begin{align}
\mathcal{A}_{1}^{2}\left(u,v\right) & =\frac{1}{n}\sum_{i=1}^{n}\left[\frac{i}{n}-p\left(d^{\mathsf{A}}\right)-\frac{1}{2n}\right]d_{(i)}^{\mathsf{A}}\\
 & =\frac{1}{n}\sum_{i=1}^{n}\left[\frac{i}{n}-p\left(d^{\mathsf{A'}}\right)-\frac{1}{2n}\right]d_{(i)}^{\mathsf{A}}+\left[p\left(d^{\mathsf{A'}}\right)-p\left(d^{\mathsf{A}}\right)\right]\frac{1}{n}\sum_{i=1}^{n}d_{(i)}^{\mathsf{A}}\\
 & =\mathcal{A}_{1}^{2}\left(u',v\right)+\frac{1}{n}\sum_{i=1}^{n}\left[\frac{i}{n}-p\left(d^{\mathsf{A'}}\right)-\frac{1}{2n}\right]\left[d_{(i)}^{\mathsf{A}}-d_{(i)}^{\mathsf{A'}}\right]+[p\left(d^{\mathsf{A'}}\right)-p\left(d^{\mathsf{A}}\right)](\mu_{v}-\mu_{u})\\
 & =\underset{\mathrm{structural}}{\underbrace{\mathcal{A}_{1}^{2}\left(u',v\right)}}+\underset{\mathrm{exchange}}{\underbrace{\frac{1}{n}\sum_{i=1}^{n}a_{i}^{\mathsf{A'}}\left[d_{(i)}^{\mathsf{A}}-d_{(i)}^{\mathsf{A'}}\right]}}+\underset{\mathrm{growth}}{\underbrace{\left[p\left(d^{\mathsf{A'}}\right)-p\left(d^{\mathsf{A}}\right)\right]\left[\mu_{v}-\mu_{u}\right]}}
\end{align}
\textbf{Scale-independent mobility index}: with $d_{i}^{\mathsf{S}}=\frac{v_{i}}{\mu_{v}}-\frac{u_{i}}{\mu_{u}}$
and $d_{i}^{\mathsf{S'}}=\frac{v_{i}}{\mu_{v}}-\frac{u'_{i}}{\mu_{u}}$,
we have 
\begin{align}
\mathcal{S}_{1}^{2}\left(u,v\right) & =\frac{1}{n}\sum_{i=1}^{n}\left[\frac{i}{n}-p\left(d^{\mathsf{S}}\right)-\frac{1}{2n}\right]d_{(i)}^{\mathsf{S}}\ \\
 & =\frac{1}{n}\sum_{i=1}^{n}\left[\frac{i}{n}-p\left(d^{\mathsf{S'}}\right)-\frac{1}{2n}\right]d_{(i)}^{\mathsf{S}}+[p\left(d^{\mathsf{S'}}\right)-p\left(d^{\mathsf{S}}\right)]\underset{=0}{\underbrace{\frac{1}{n}\sum_{i=1}^{n}d_{(i)}^{\mathsf{S}}}}\\
 & =\frac{1}{n}\sum_{i=1}^{n}\left[\frac{i}{n}-p\left(d^{\mathsf{S'}}\right)-\frac{1}{2n}\right]d_{(i)}^{\mathsf{S'}}+\frac{1}{n}\sum_{i=1}^{n}\left[\frac{i}{n}-p\left(d^{\mathsf{S'}}\right)-\frac{1}{2n}\right]\left(d_{(i)}^{\mathsf{S}}-d_{(i)}^{\mathsf{S'}}\right)\\
 & =\underset{\mathrm{structural}}{\underbrace{\mathcal{S}_{1}^{2}\left(u',v\right)}}+\underset{\mathrm{exchange}}{\underbrace{\frac{1}{n}\sum_{i=1}^{n}a_{i}^{\mathsf{S'}}\left[d_{(i)}^{\mathsf{S}}-d_{(i)}^{\mathsf{S'}}\right]}}
\end{align}
\textbf{Translation-independent mobility index}: with $d_{i}^{\mathsf{T}}=[v_{i}-u_{i}]-[\mu_{v}-\mu_{u}]$
and $d_{i}^{\mathsf{T'}}=[v_{i}-u'_{i}]-[\mu_{v}-\mu_{u'}]$, we have
\begin{align}
\mathcal{T}_{1}^{2}\left(u,v\right) & =\frac{1}{n}\sum_{i=1}^{n}\left[\frac{i}{n}-p\left(d^{\mathsf{T}}\right)-\frac{1}{2n}\right]d_{(i)}^{\mathsf{T}}\ \\
 & =\frac{1}{n}\sum_{i=1}^{n}\left[\frac{i}{n}-p\left(d^{\mathsf{T}}\right)+p\left(d^{\mathsf{T'}}\right)-p\left(d^{\mathsf{T'}}\right)-\frac{1}{2n}\right]d_{(i)}^{\mathsf{T}}\\
 & =\frac{1}{n}\sum_{i=1}^{n}\left[\frac{i}{n}-p\left(d^{\mathsf{T'}}\right)-\frac{1}{2n}\right]d_{(i)}^{\mathsf{T}}+[p\left(d^{\mathsf{T'}}\right)-p\left(d^{\mathsf{T}}\right)]\underset{=0}{\underbrace{\frac{1}{n}\sum_{i=1}^{n}d_{(i)}^{\mathsf{T}}}}\\
 & =\frac{1}{n}\sum_{i=1}^{n}\left[\frac{i}{n}-p\left(d^{\mathsf{T'}}\right)-\frac{1}{2n}\right]\left(d_{(i)}^{\mathsf{T}}+d_{(i)}^{\mathsf{T'}}-d_{(i)}^{\mathsf{T'}}\right)\\
 & =\frac{1}{n}\sum_{i=1}^{n}\left[\frac{i}{n}-p\left(d^{\mathsf{T'}}\right)-\frac{1}{2n}\right]d_{(i)}^{\mathsf{T'}}+\frac{1}{n}\sum_{i=1}^{n}\left[\frac{i}{n}-p\left(d^{\mathsf{T'}}\right)-\frac{1}{2n}\right]\left(d_{(i)}^{\mathsf{T}}-d_{(i)}^{\mathsf{T'}}\right)\\
 & =\underset{\mathrm{structural}}{\underbrace{\mathcal{T}_{1}^{2}\left(u',v\right)}}+\underset{\mathrm{exchange}}{\underbrace{\frac{1}{n}\sum_{i=1}^{n}a_{i}^{\mathsf{T'}}\left[d_{(i)}^{\mathsf{T}}-d_{(i)}^{\mathsf{T'}}\right]}}
\end{align}

\end{document}